\documentclass[preprints,article,accept,moreauthors,pdftex]{Definitions/mdpi}

\firstpage{1} 
\pubvolume{xx}
\issuenum{1}
\articlenumber{5}
\pubyear{2019}
\copyrightyear{2019}
\history{Received: date; Accepted: date; Published: date}

\Title{Lapsing Quickly into Fatalism: Bell on Backward Causation} 

\Author{Travis Norsen$^{1,\ddagger}$ and Huw Price$^{2,\dagger,\ddagger}$
}

\AuthorNames{Firstname Lastname, Firstname Lastname and Firstname Lastname}

\address[1]{%
$^{1}$ \quad Smith College; tnorsen@smith.edu\\$^{2}$ \quad University of Cambridge; hp331@cam.ac.uk}

\corres{Correspondence: hp331@cam.ac.uk 
}

\firstnote{Current address: Trinity College, Cambridge CB2 1TQ, UK} 
\secondnote{These authors contributed equally to this work.}

\abstract{This is a dialogue between Huw Price and Travis Norsen,  loosely inspired by a letter that Price received from J.~S.~Bell in 1988. The main topic of  discussion is Bell's views about retrocausal approaches to quantum theory, and their relevance to contemporary issues.}


\renewcommand{\textit}{\emph}
\begin{document}


\section{Introduction}

This paper is a dialogue between Huw Price and Travis Norsen, loosely inspired by a letter that Price received from John Bell in 1988. The main topic of  discussion is Bell's views about retrocausal approaches to quantum theory, and their relevance to contemporary issues.

 
\section{Price (I)}
As far as I can recall, I first heard about Bell’s Theorem at a workshop at Wolfson College, Oxford, in the Spring of 1977. (I was in Oxford that year as an MSc student in Mathematics.) I remember little of the talk, except that the speaker noted in passing that Bell's argument required the assumption that the properties of the particles concerned didn’t depend on the future measurement settings. At any rate, that's what I took him to be saying. I certainly don't recall the exact words, and we'll see in a moment that there's another possibility for what he might have meant -- but that was the sense of the assumption I took away. 

I remember even this much because I was  puzzled at the time that this assumption seemed to be regarded as uncontroversial.  I had read some philosophy of time by that point -- enough to be convinced that past and future are equally real, and to be familiar with the idea that time-asymmetry in the physical world is a statistical matter.  Yet here was a time-asymmetric assumption about what can affect what -- on the face of it, not a statistical matter --  playing a crucial role in Bell's argument. Everyone seemed to agree that the argument  led to a highly counterintuitive conclusion. But the option of avoiding the conclusion by rejecting the assumption  didn’t seem to be  on the table. 

More than forty years later, I’m still puzzled. I've returned to the issue at intervals over those years, always looking for a good reason for closing what seemed to me an open door – a door, among other things, to a potential resolution of the tension between quantum theory and special relativity. To me, Nature seemed to be offering us a huge hint, a hint revealed in Bell's work, that our intuitions about what can depend on what are unreliable in the quantum world. Yet few of my growing circle of friends and colleagues who knew about these issues -- who knew a great deal more than I did, in most cases -- ever seemed able to hear the hint. I often wondered what I was missing.

After my year in Mathematics at Oxford I shifted to Philosophy in Cambridge. There, as some sort of sideline to my main thesis project on probability, I spent some time on the puzzle I had brought with me from Oxford. In \cite{Price78}, a piece written in November 1978, I discuss what is in effect the following assumption, central to what is now called the \textit{ontological models framework} \cite{Harrigan10}: 
\begin{quote}
In the ontological models framework, it is assumed that the probability measure representing a quantum state is independent of the choice of future measurement setting. (Leifer \cite[140]{Leifer14})
\end{quote}
My 1978 piece argues that this kind of assumption is `very difficult to justify on metaphysical grounds', and notes that abandoning it has a very interesting potential payoff, given its crucial role in the no-go theorems of Bell and of Kochen \& Specker. But, as the present ubiquity of the ontological models framework demonstrates,   this hasn't become a common concern. Now, as in the 1970s, the assumption in question usually passes without comment -- it is simply part of the model.


\begin{figure}[t!]
\centering
\includegraphics[width=11cm]{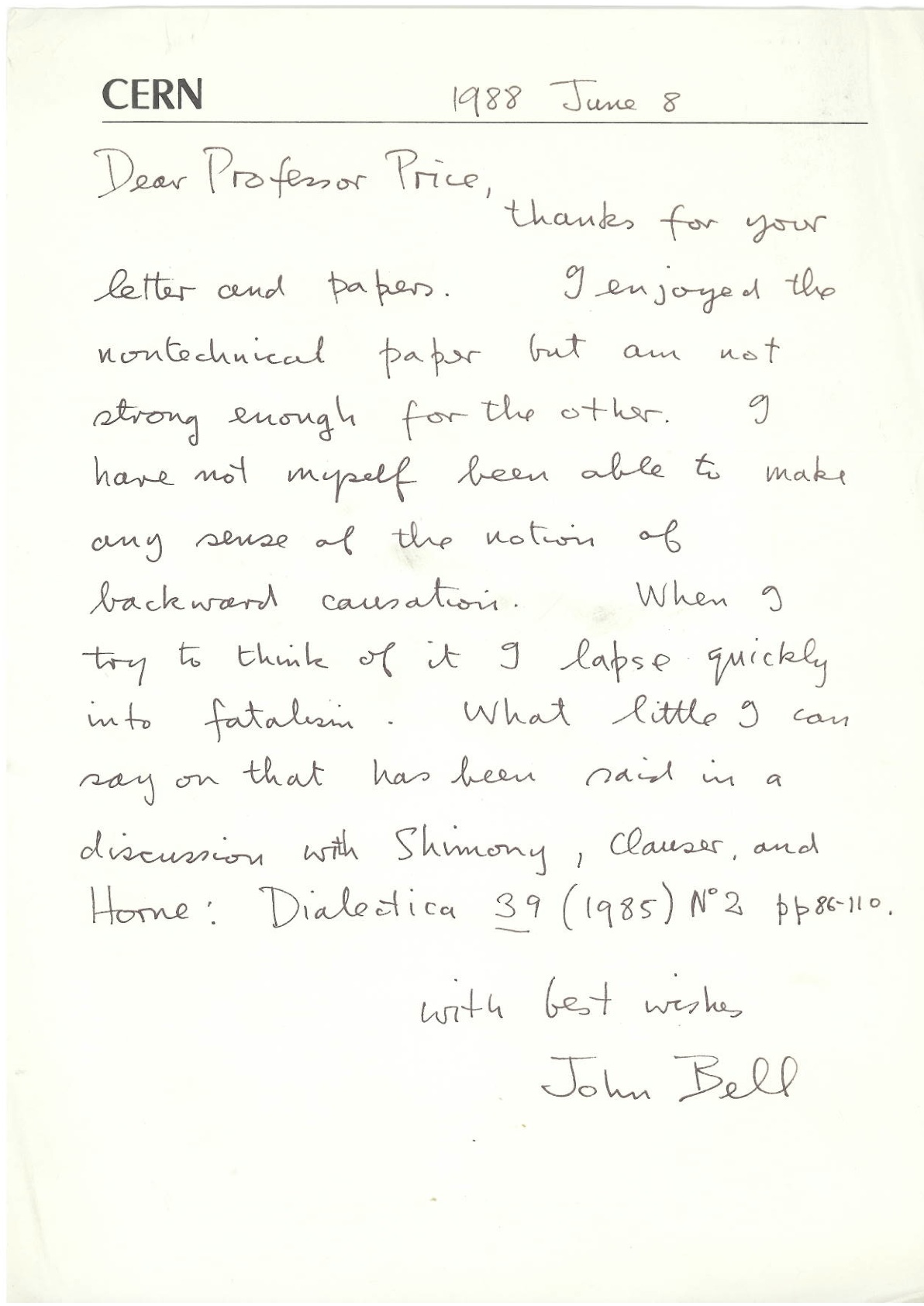}
\caption{The letter from Bell.}
\end{figure}  


A couple of years later again, now a postdoc at ANU, Canberra, I worked on this little obsession some more. I focussed on the work of the Oxford philosopher Michael Dummett (who had been at the Wolfson workshop in 1977, I believe). Dummett had two well-known papers  defending the coherence of retrocausality \cite{Dummett54, Dummett64}, and in a piece published in \textit{Synthese} in 1984 \cite{Price84} I offered some refinements to Dummett's arguments, and noted their potential application in the quantum case.

Back at ANU later in 1980s, I wrote the early drafts of a piece that eventually appeared in \textit{Mind} in 1994 \cite{Price94}. I think it was a draft of this piece, together with my \textit{Synthese} piece \cite{Price84} from 1984, that I sent to John Bell in 1988. His brief reply appears as Figure 1. About the idea of retrocausality, he says this:
\begin{quote}
 I have not myself been able to make any sense of the notion of backward causation. When I try to think of it I lapse quickly into fatalism.   
\end{quote}
 For `what little I can say', Bell then refers to a published discussion \cite{Bell85} with Shimony, Clauser and Horne.

Bell's reference to fatalism in this letter certainly chimed with some of his published remarks, to the effect that to abandon the assumption in question, we would have to abandon the assumption that we are free to choose the measurement settings. This is from the piece referred to in his letter, for example: 
  \begin{quote}
     It has been assumed that the settings of instruments are in some sense free variables — say at the whim of the experimenters -- or in any case not determined in the overlap of the backward lightcones. Indeed without such freedom I would not know how to formulate any idea of local causality, even the modest human one \cite{Bell85}.
 \end{quote}
In the light of such remarks, the option of abandoning this assumption became known as the \textit{free will,} or \textit{freedom of choice,} loophole in Bell's Theorem. 

As I'll explain, Bell's letter didn't do much to remove my sense of puzzlement. But it did help me to see that there are two very different models for what would be involved in abandoning the crucial assumption in Bell's argument -- and that it was at least unclear whether Bell himself had  properly distinguished them. Moreover, the model discussed in  \cite{Bell85} was not the one that had interested me in the first place. 
 
Accordingly, when I next wrote about these topics \cite{Price95,Price96}, I tried to distinguish the two models. Briefly, the difference is this. Both models involve correlations between measurement settings and properties of an incoming particle. In other words, both reject the assumption often referred to as \textit{Statistical Independence.}  But they propose to explain this correlation in very different ways.

 It is a familiar idea that we need to distinguish correlation from causation, and that the same pattern on correlations may be compatible with more than one causal explanation. In susceptible folk, for example, eating chocolate is said to be correlated with onset of migraine, a short time later. If so, this might be because chocolate \textit{causes} migraine, or because migraine and a craving for chocolate are both \textit{effects} of some underlying physiological cause (a \textit{common cause,} as causal modellers say). 
 
 In the discussion in \cite{Bell85}, it is assumed that the explanation of a correlation between measurement settings and underlying particle properties would have to be of the latter kind. In other words, it would have to be due to some common cause in the overlap of the past lightcones of the setting and the particle. Call this the \textit{Common Past Hypothesis} (CPH). In Figure 2, adapted from a famous diagram due to Bell himself, the  common cause is shown affecting both the measurement setting \textit{a} on the left, and `beables' $\lambda$ on the right (which in turn can affect the measurement outcome \textit{B}) 

\begin{figure}[t!]
\centering
\includegraphics[width=14cm]{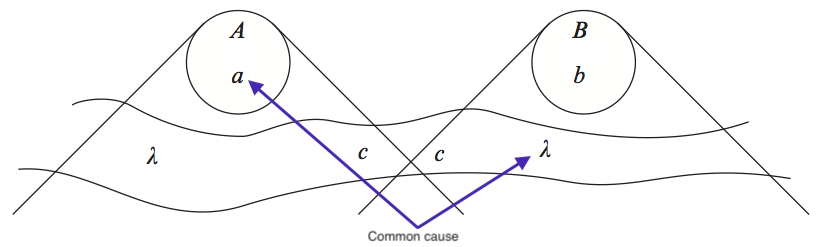}
\caption{The Common Past Hypothesis (CPH)}\vspace{0.5cm}
\end{figure} 

\begin{figure}[t!]
\centering
\includegraphics[width=14cm]{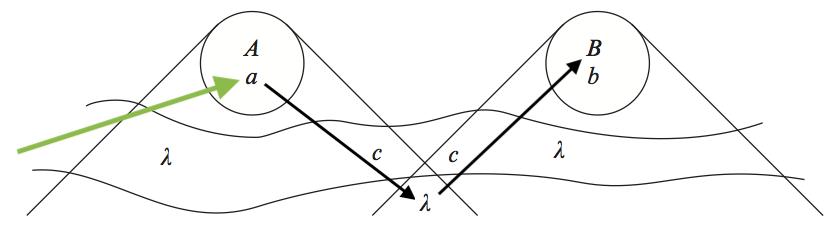}
\caption{The Common Future Hypothesis (CFH)}
\end{figure}

 In the case that I had in mind, however, the causal structure was different. As with chocolate causing migraine, the  correlation was  to be explained by the hypothesis that the properties of the incoming particle were themselves causally influenced by the measurement settings. Of course, unlike in the chocolate and migraine case, this causal influence needed to work from future to past – it was \textit{retrocausality,} as we now say. Setting aside for the moment the unfamiliarity of that idea -- I am aware that some readers will see it as the elephant in the ointment for this proposal! -- the difference between this explanation of the correlations and the one involving a common cause is as stark as in the migraine case. They are completely different hypotheses.
 
Normally, of course, we take the measurement settings themselves to have causes in their own past, namely the choices of human experimenters, or the devices to which we humans delegate control. Adding this to the picture, we get the kind of causal model depicted in Figure 3. The green arrow from the left represents the experimenter's choice of the measurement setting \textit{a}, and the black arrows represent the effect of this choice on $\lambda$ and hence on \textit{B}. A similar path would have to operate from right to left, of course. Note that nothing hangs on the fact that the green arrow is shown acting from outside the past lightcone. (We'll see that it is relevant that it comes from `outside' in a different sense -- it is an experimental intervention, of a kind universal in science.) 
 
 In this second model the correlation between measurement choices and earlier particle properties is explained \textit{in the future,} by the fact that the particle encounters the measurement device there -- a device whose setting, in turn, is determined by an earlier choice on the part of an experimenter. Accordingly, let's call this the \textit{Common Future Hypothesis} (CFH).

It was easy to see how both CPH and CFH might lead someone to fatalism. However, it is crucial to see that they do so by \textit{very different routes.} In the case of CPH, measurement settings are treated as \textit{effects} of the postulated common cause. In the language of the causal modelling framework, the measurement settings are therefore being treated as \textit{endogenous} variables (meaning, as Hitchcock puts it, `that their values are determined by other variables in the model' \cite{Hitchcock20}). Here, already, we have a stark contrast with the normal status of experimental settings in scientific models. Normally these are exogenous, `meaning that their values are determined outside of the system' \cite{Hitchcock20}. It is easy to see how this change of status might seem incompatible with a very down-to-earth sense of experimental freedom -- and might seem, as Wiseman  puts it, to `undercut the core assumptions necessary to undertake scientific experiments' \cite{Wiseman2006}. 

This objection goes back at least to Bell's discussion with Shimony, Clauser and Horne in \cite{Bell85} (an exchange originally published in 1976). As Shimony, Clauser and Horne put it:
\begin{quote}
  In any scientific experiment in which two or more variables are supposed to be randomly selected, one can always conjecture that some factor in the overlap of the backwards light cones has controlled the
presumably random choices. But, we maintain, skepticism of this sort
will essentially dismiss all results of scientific experimentation. Unless
we proceed under the assumption that hidden conspiracies of this sort
do not occur, we have abandoned in advance the whole enterprise of
discovering the laws of nature by experimentation.  
\end{quote}
This challenge to CPH seems to me to be entirely correct. As I say, however, what I had in mind was CFH. From the beginning at Wolfson College in 1977, I had taken the interesting and questionable assumption to be the claim that the properties of the particles could not \textit{depend on} future measurement settings. CPH doesn't challenge that assumption at all. It simply makes both the measurement settings and the particle properties depend on some third thing in their common past. 

This means that I have not been bothered by these objections to CPH. They don't trouble CFH, where there is no bar to treating measurement settings as exogenous variables. (That's the sense in which the green arrow in Figure 3 `comes from outside'.) In CFH the threat of fatalism comes from the fact that the particle 'already knows' the choice of measurement setting, before it is made. In the case of an observation on a photon from a distant galaxy, for example, the model requires that the particle has `known' for billions of years what the measurement setting would be. Isn't that incompatible with the ordinary belief that we have a free choice in the matter?

At this point I felt there were two things to say. First, this argument is {logically} identical to an ancient argument for fatalism, an argument starting from the assumption that statements about the future have determinate truth values. Theologians once worried about problems in this vicinity. Doesn't God's knowledge of the future deprive us of free will, for example? Leibniz  discusses such issues, noting that ‘the sophism which ends in a decision to trouble oneself over nothing will haply be useful sometimes to induce certain people to face danger fearlessly’ \cite[153]{Leibniz51}. In the following instance the strategic fatalist  is Henry V, admonishing Westmoreland for wishing for reinforcements for their impending battle at Agincourt:
\begin{verse}
No, my fair cousin:\\
If we are mark’d to die, we are enow\\
To do our country loss; and if to live,\\
The fewer men the greater share of honour. (Henry V, Act IV, Sc.~III)
\end{verse}
If the concern about fatalism in CFH amounts to no more than this, it is a very damp squib indeed. Physicists shouldn't allow themselves to be mugged by medieval kings and theologians. 

Second, and more interestingly, the concern about CFH might rest on issues about causal loops. Suppose we could \textit{find out} the relevant property of the incoming particle, before it reaches the measuring device, and use the information to change the measurement setting. Contradiction? This is a reasonable concern, but as I  noted already in \cite{Price84}, Dummett's work identifies the solution. Backward causation is safe from such concerns, so long as it is impossible to `find out' about the effect in question, before the choice of the future setting on which it depends. The restrictions that quantum theory puts on measurement seemed to offer a prospect that it could exploit Dummett's loophole.

For these reasons, I felt that in the case of CFH, concerns about fatalism were unwarranted. This left me eventually with the following view about Bell's position, extrapolated from his letter. Either Bell hadn't sufficiently distinguished CPH and CFH in his own thinking, and was assuming that the (valid) concern about the former would also encompass that latter. Or he had distinguished them, and was relying on a much more questionable argument for fatalism in the latter case. This is what I meant when I said that Bell's letter didn't resolve my sense of puzzlement. Perhaps there was some some third option that I was missing, but I couldn't see what it would be.

It would be nice to be able to report that these issues are more clearly understood these days, but  confusion persists in some quarters. In particular, it still seems to be widely believed that ruling out CPH would be sufficient to close the loophole in Bell's Theorem associated with Statistical Independence. These quotations are from two recent pieces (\cite{Giustina15},\cite{Handsteiner17}) by Anton Zeilinger and collaborators, for example:

\begin{quote}
  The \textit{freedom-of-choice loophole} refers to the requirement, formulated by Bell, that the setting choices are ``free or random'' \cite[232]{Bell04}. For instance, this would prohibit a possible
interdependence between the choice of measurement settings and the properties of the system being measured.  
\end{quote}

\begin{quote}
    A ... major loophole, known variously as the
freedom-of-choice, measurement-independence, or
setting-independence loophole ..., concerns the
choice of measurement settings. In particular, the
derivation of Bell’s inequality explicitly assumes that
there is no statistical correlation between the choices of
measurement settings and anything else that causally
affects both measurement outcomes. Bell himself
observed forty years ago that, ``It has been assumed
that the settings of instruments are in some sense free
variables---say at the whim of experimenters---or in any
case not determined in the overlap of the backward light
cones''. 
\end{quote}
In both cases here, it is simply taken for granted that the only avenue for statistical dependence between measurement settings and `anything else that causally affects both measurement outcomes' is one that involves some common cause, acting in the past. (The second paper claims to be `pushing back by \textasciitilde 600 years the most recent time by which any local-realist influences could have engineered the observed Bell violation'.)

I don't want to overstate this claim of confusion. Some recent writers are admirably clear that CFH, or retrocausality, is a distinct proposal for rejecting Statistical Independence. (Leifer \cite{Leifer14} links it to the option of rejecting the ontological models framework; see \cite{FriedrichEvans19,WhartonArgaman20} for recent surveys.) Moreover, we now have a better understanding of the issue of the relationship between time-symmetry and retrocausality, another factor in my interest in the case from the beginning. My own result \cite{Price12}, extended and generalised by  Leifer and Pusey \cite{LeiferPusey17, Leifer17}, shows how in quantum theory time-symmetry may require retrocausality, for a reason not present in the classical case. As Leifer and Pusey note, the argument depends on thinking about time-like analogues of EPR-Bell arguments: the EPR argument in my case, and Bell's extension in theirs.

For me, the effect of this recent work has been to increase my sense that Nature is offering us a loud hint in Bell's results, a hint to which many people interested in these topics are curiously deaf. (If anything, I now feel that there are two hints, one from Lorentz invariance and one from time-symmetry.) As always, however, I'm conscious that there may be objections of principle that I cannot see, and that Bell did see, perhaps. If there's any trace of such objections in Bell's letter from 1988, I'm hoping that this discussion will unearth it. I'm also conscious that I haven't yet touched on the idea of so-called `superdeterminism', or the suggestion that something objectionably `conspiratorial' is required for a proposal such as CFH. Again, if there is a valid objection to CFH of this kind, I'm hoping this discussion will highlight and clarify it.

Before I yield the floor to m'colleague, let me mention again the (supposed) elephant in the ointment for CFH, the idea of backward causation itself. As I noted above, my own early discussion of these ideas took a deliberate path via some classic philosophical work on the issue, that of Michael Dummett \cite{Dummett54,Dummett64}. Since those early days, I have written extensively on the issue of the direction of causation \cite{Price96,PriceWeslake10}. In those investigations, I haven't yet found anything that ought to count as a fly in the ointment for CFH, much less an elephant. On the contrary, Dummett's identification of a loophole in objections based on causal loops remains precisely the fly-free balm that CFH requires, in my view. 

I know of one place where Bell himself discusses backward causation explicitly. In `La Nouvelle Cuisine', seeking a characterisation of the sense of locality apparently implied by special relativity, he suggests that it might be defined in terms of cause and effect \cite[235]{Bell04}:
\begin{quote}
  As far as I know, this was first argued by Einstein, in the context of special relativity theory. In 1907 he pointed out that if an effect followed its cause sooner than light could propagate from the one place to the other, then in some other inertial frames of reference the ‘effect’ would come before the ‘cause’! He wrote
\begin{quote}
  \ldots\ in my opinion, regarded as pure logic \ldots\ it contains no contradictions; however it absolutely clashes with the character of our total experience, and in this way is proved the impossibility of the hypothesis \ldots\ of a causal chain going faster than light.  
\end{quote}  
\end{quote}
Bell goes on to explain what Einstein had in mind -- a case of causal loops, exploiting the fact that the effects precede their causes in some inertial frames. Perhaps this is evidence that Bell himself thought that backward causation is excluded by these causal loop arguments. If so, then the escape hatch he needed is close at hand. All such arguments depend on the assumption that backward causation  could be used to signal (to tell one's grandmother to avoid her unhappy marriage to grandfather, perhaps, in the classic paradox). But as Bell himself made clear, the Bell correlations imply causality without signalling. So long as retrocausality stays on the same side of the line, it is safe from paradox. (Once again, this is Dummett's loophole, in effect.) 

As for Einstein's own attitude to retrocausality, this little fragment suggests an admirable and characteristically empirical attitude. The objection is that `it absolutely clashes with the character of our total experience'. Of course, Einstein didn't know about the `aspect' of our total experience that was to be revealed by Bell's work, and by the experiments it inspired. So it would be presumptuous, to say the least, to read into Einstein's remarks any general prohibition on retrocausal models.

\section{Norsen (I)}

Thanks to Huw for inviting me to join him in what promises to be an illuminating dialogue and for perfectly setting the context with that beautiful opening statement. 

Since it does not involve any direct interaction with (or letters from) Bell, my own personal biography vis-a-vis Bell’s theorem is far less interesting than Huw’s.  But in so far as I can reconstruct it, I can report that I first learned of Bell’s theorem from David Albert’s book \emph{Quantum Mechanics and Experience} \cite{albert}, which I stumbled across at Orca Books in downtown Olympia, Washington (USA) during the winter vacation in the middle of my sophomore year of college, which would have been late 1994 or early 1995 (i.e., about four years after Bell’s untimely death).  Albert’s book (and the many other books and articles on the foundations of QM that I subsequently began to devour) made a significant impression on me and gave me a youthful confidence to raise skeptical questions about orthodox quantum mechanics as it was presented in my physics courses.  

Given what I now know of the typical attitudes toward quantum foundations in the physics community, it is rather surprising that my undergraduate professors during this period were so incredibly supportive of my interests in unorthodox viewpoints.  But they must have found my passion at least a little naive, and, in hindsight, I can now see that they were not completely wrong.  

Huw remembers questioning a very specific and subtle assumption in Bell’s theorem the first time he encountered it.  For me the story is very different.  I spent years knowing that Bell’s theorem was crucially important, simply because everybody unanimously agreed that it was, but having basically no clear idea at all what to make of it, because different authors all seemed to present their own totally unique version of the theorem’s logical structure.  David Albert had claimed that Bell’s theorem proved that nature was non-local, but other commentators told very different stories, usually along the lines of:  Bell had refuted determinism, or the related and EPR-inspired hidden-variables program, and had therefore put the final nail in Einstein’s coffin and proved once and for all that the orthodox interpretation of Bohr and Heisenberg was the only viable one.  

Some semblance of clarity only emerged while I was in graduate school (nominally pursuing a PhD in theoretical nuclear astrophysics but really spending at least half of my time secretly studying quantum foundations), when it finally occurred to me that perhaps Bell himself might have an illuminating perspective on his own theorem.  Reading Bell’s collected papers \cite{Bell04} turned out to be very helpful indeed.  Even his more technical papers were completely accessible and clear, and Bell’s ability to explain his reasoning, crisply and cleverly, was truly masterful.  I went from having no idea what Bell’s theorem actually proved, to having no idea how any controversy could remain when Bell had laid everything out so perfectly.

Huw’s opening statement focused mostly on the notion of \emph{Statistical Independence}, which is the main thing we’ll be talking about.  But just to set the context and pre-empt any possible misunderstanding or miscommunication, I think it will be helpful to step back and lay out the overall structure of Bell’s theorem (at least as Bell himself understood it and helped me to understand it).  

There are two assumptions.  One is the \emph{Statistical Independence} that Huw has discussed, according to which (in the usual EPR-Bell sort of setup) the measurement settings \emph{a} and \emph{b} are ``free'' or ``exogenous'' and therefore (at least as long as we set aside the idea of retro-causation that Huw wants us to consider) uncorrelated with the variables $\lambda$ which characterize the physical state of the particle pair at some earlier time.  There will be much more to say about this assumption as the discussion proceeds.

But I wanted to make sure to acknowledge explicitly that there is also another assumption -- which Bell and I and most others would consider in some sense the more central and important assumption -- namely, \emph{Local Causality}.  Bell’s careful and important mathematical formulation of this notion is intended to capture the qualitative idea, motivated by relativity, that the causal influences \emph{on} a given event are to be found exclusively in that event’s past light cone, and the causal influences \emph{of} a given event (on other events) are to be found exclusively in the future light cone.  The idea, in short, is that causal influences propagate (from the past toward the future!) always at the speed of light or slower.  Readers unfamiliar with Bell’s formulation of \emph{Local Causality} are urged to read his final and clearest presentation in ``La Nouvelle Cuisine'' \cite[232]{Bell04} and/or my own dissection in ``J.S. Bell’s Concept of Local Causality'' \cite{norsen}.  

As a brief aside that will be relevant later, let me stress here that the core virtue of Bell’s formulation of \emph{Local Causality} is that it is totally and completely generic.  The formulation is not in terms of the proprietary concepts (e.g., quantum mechanical wave functions) of some specific candidate theory, but is exclusively in terms of the un-sectarian notion that Bell invented for the purpose:  ``beables'', which simply means whatever some candidate theory posits to exist.  And (unlike for example the conditions known as ``Parameter Independence'' and ``Outcome Independence'') Bell’s \emph{Local Causality} does not make reference to any specific type of process or situation and in particular does not imply or require any sub-classification of beables into distinct sub-types (e.g., those which are human-controllable ``parameters'' vs. those which are uncontrollable ``outcomes'').  

Bell’s concept of \emph{Local Causality}, that is, possesses the same virtues that Bell demanded of candidate theories when he complained that orthodox quantum theory, with its special \emph{ad hoc} rules for how systems behave during measurements, was ``unprofessionally vague and ambiguous'' \cite[173]{Bell04}. As he elaborated elsewhere, terms such as \emph{measurement}, \emph{observable}, \emph{system}, and \emph{apparatus} ``...however legitimate and necessary [they might be] in application, have no place in a \emph{formulation} with any pretension to physical precision'' \cite[215]{Bell04}.  To avoid suffering from the sort of ``measurement problem'' that afflicts orthodox quantum theory, Bell thought, the ontological posits and dynamical laws of a proper candidate fundamental theory should be stated in precise mathematical way, without vague, anthropocentric terms or distinctions.  Bell thus appreciated the professionalism of various unorthodox formulations of quantum theory such as the pilot-wave theory, spontaneous collapse theory, and (to a lesser extent) Everett’s many-worlds theory... and we should appreciate the professionalism  of Bell’s \emph{Local Causality} on similar grounds.

Anyway, returning to Bell's theorem, the two assumptions, \emph{Local Causality} and \emph{Statistical Independence}, turn out to jointly entail something (“Bell’s inequality”) which I think Huw and I will agree is now known, from experiment, to be false.  So at least one of the two assumptions -- \emph{Local Causality} and \emph{Statistical Independence} -- must be rejected.  

Bell’s view, and the view that I and many other commentators have tended to adopt, is that \emph{Statistical Independence} is something like an unquestionable assumption of empirical science, the denial of which amounts to endorsing a kind of cosmic conspiracy theory.  Indeed, in one important earlier article on Bell’s theorem \cite{scholarpedia}, my co-authors and I called the \emph{Statistical Independence} assumption by the alternative name ``No Conspiracies'' -- much, no doubt, to the annoyance of Huw and others of his ilk!  

So as part of this opening statement let me just stipulate for the record that Huw is entirely correct to insist that the idea of backwards-in-time causation -- the Common Future Hypothesis, CFH -- potentially provides (or at least appears to potentially provide) a non-conspiratorial way of violating the \emph{Statistical Independence} assumption.  So this very much deserves to be peeled apart and examined carefully, hence my excitement to participate in this dialogue.  

But I have a big-picture question that I think should be addressed here at the outset.  The people who want to deny the \emph{Statistical Independence} assumption -- not on the basis of retrocausation and the CFH, but rather on the basis of the CPH and hence what Huw and I would agree is a scientifically-unacceptable kind of conspiracy -- want to do so \emph{in order to save Local Causality}.  That is, the end-game of the conspiracy theorists is to find a way of reconciling the relativity-based idea that causal influences propagate (from the past toward the future!) always at the speed of light or slower, with the empirical violation of Bell's inequality.  

But this cannot be your endgame, Huw, since (as I have tried to stress) the other premise of Bell's theorem, \emph{Local Causality}, also has a (not merely statistical) arrow of time built into it.  We could put the point like this.  The kind of retro-causation that you want to use to provide a non-conspiratorial ground for rejecting \emph{Statistical Independence}, also just blatantly and openly violates Bell’s notion of \emph{Local Causality}; it says, after all, that the causal influences \emph{on} certain events are to be found in their future light cones!  So apparently, despite both focusing skeptical attention on the \emph{Statistical Independence} assumption, you are not at all trying to achieve the same thing as the conspiracy theorists.  So what exactly are you trying to achieve?  

I think I have a sense of what your answer will be, but I’m sure it will help focus the subsequent discussion to have this laid out explicitly.

\section{Price (II)}

Travis mentions a fateful Olympian encounter with Albert’s excellent book, \textit{Quantum Mechanics and Experience,} that led him into quantum foundations. This gives me an opportunity to recommend Travis's own recent text, \textit{Foundations of Quantum Mechanics.} Contemporary versions of the 1990s Travis, or the 1970s me, would be just as lucky to encounter this book as he was to encounter Albert's -- in some ways, even more so. Among other things, Travis's book, unlike Albert's, offers a rich sense of engagement with the founders of quantum foundations -- Einstein, Bell, and many others. Travis achieves this by working in well-chosen words from these greats, and I'll draw on some of those in a moment.

Travis points out a contrast between retrocausalists such as me, who look for violation of Statistical Independence via what I have called CFH, and those we are now labelling conspiracy theorists, who do so via CPH. As Travis says, proponents of CPH are trying to save Bell's principle Local Causality. That can't be my goal, because, as he puts it, Local Causality has an ``arrow of time built into it''. So what exactly am I trying to achieve?
It's a good question. The main part of my answer is that I want to defend a more basic sense of Locality, and show how the world might violate Bell's Local Causality but respect the more basic notion. 

What is the more basic notion? Here it is in words from Einstein, quoted by Bell, in a passage reproduced in Travis's book: 
\begin{quote}
    The following idea characterizes the relative independence of objects far apart in space (A and B): external influence on A has no direct influence on B. \cite[109]{Norsen17}
\end{quote}
Call this \textit{Einstein Locality.} Retrocausal models want to preserve Einstein Locality at the cost of Bell's Local Causality -- at the same time explaining why this isn't much of a cost at all, once we understand the limitations of Bell's version. The major limitation is the way in which \textit{Local Causality} simply builds in a causal arrow of time. Einstein Locality says nothing about time -- we could replace `far apart in space' with `far apart in spacetime, in any direction' and still have much the same idea.

Travis describes the ideas that Bell's \textit{Local Causality} is intended to capture like this::

\begin{quote}
    Bell’s \ldots\ mathematical formulation \ldots\ is intended to capture the qualitative idea, motivated by relativity, that the causal influences on
a given event are to be found exclusively in that event’s past light cone, and the causal influences of a
given event \ldots\ are to be found exclusively in the future light cone. The idea, in short,
is that causal influences propagate (from the past toward the future!) always at the speed of light or
slower. 
\end{quote}
There are two parts to this qualitative idea, one the restriction to lightcones motivated by relativity, and other -- much older, obviously -- that causal influences propagate from past to future. Let's call the latter the \textit{Causal Arrow of Time,} or CAT for short. Note that CAT actually combines two things, first a distinction between cause and effect, and second the claim that the cause--effect `arrow' lines up with the earlier--later `arrow'. (Unless the causal relation is itself asymmetric, in makes no sense to say that it points in a particular direction.)

A generation  before Schr\"odinger's famous feline, this CAT played a role in the most notorious rejection of causation in modern philosophy. In 1912 Bertrand Russell argued that modern physics had no use for causation, and that philosophy should therefore discard it too: ``The law of causation,'' Russell said, ``Like much that passes muster among philosophers, is a relic of a bygone age, surviving, like the monarchy, only because it is erroneously supposed to do no harm.'' \cite{Russell13} One of Russell's main arguments is that there is no asymmetric relation in physics that we could identify with causation. The computer scientist Judea Pearl, himself a leading contemporary writer on causation, sums up Russell's point like this: ``[T]he laws of physics are all symmetrical, going both ways, while causal relations are unidirectional, going from cause to effect.'' \cite{Pearl09}

Russell's argument had little practical effect, and indeed, as Patrick Suppes pointed out later \cite{Suppes70}, physicists themselves often use causal notions. Still, Russell had put his finger on a puzzle -- `Russell's enigma', as Pearl calls it. As Pearl says:  
\begin{quote}
[V]ery few physicists paid attention to Russell’s enigma. They continued to write equations in the office and talk cause–effect in the cafeteria; with astonishing success they smashed the atom, invented the transistor and the laser.   \cite{Pearl09} 
\end{quote}
The long debate about nonlocality shows that it is not just in the cafeteria that these things matter in physics, but this makes the enigma all the more urgent. What is this asymmetric relation doing in physics -- or anywhere else, for that matter, in a world built on the symmetric laws of physics?

Simplifying a bit, we can distinguish three contemporary accounts of CAT (see \cite{PriceWeslake10} for discussion and references). 
\begin{enumerate}
    \item \textbf{A matter of definition.} Following Hume, we can treat CAT as a matter of definition. This view holds that the basic relations of dependence are among the symmetric relations identified by physics, and the terms `cause' and `effect' are just labels for the earlier and later of a pair of events related in this way.
    \item \textbf{Thermodynamics.} We can try to explain CAT in terms of the thermodynamic arrow of time (and in particular the so-called Past Hypothesis, or low entropy initial boundary condition). This view has been defended in recent years by writers such as Kutach, Albert, and Loewer. 
    \item \textbf{Interventionism.} The third possibility seems to originate with Frank Ramsey \cite{Ramsey29}. Ramsey, one of the fathers of the subjectivist approach to probability, takes a similar line on causation. As he puts it, ``from the situation when we are deliberating seems to \ldots\ arise the difference of cause and effect.'' In effect, Ramsey proposes an explanation of the time-asymmetry of causation (indeed, causality itself) in terms of  the epistemic perspective of agents like us. This approach has been influential in recent decades, thanks to the work of writers such as Jim Woodward, and Judea Pearl himself \cite{Woodward03, Pearl09}. It is now called \textit{Interventionism,} alluding to the central role of the idea of \textit{intervening} on a system of interest -- reaching in `from the outside', to fix the value of an exogenous variable. 
\end{enumerate}

None of these accounts of CAT seems much use to Bell, seeking to build a fundamental causal arrow into a principle for quantum foundations. The first is empty, since makes it a matter of definition that effects are later than their causes. (It would have nothing to say about probabilistic dependence between hidden variables and future measurement settings, though it would prohibit us from calling it `causality'.)  The second seems insufficiently fundamental. On the face of it, we want a theory of the quantum world that is independent of the thermodynamic environment in which a system happens to be embedded. And the third seems insufficiently  fundamental for a different reason. Its dependence on the perspective of agents like us seems deeply in tension with Bell's desire, as Travis puts it, to avoid `anthropocentric terms or distinctions'. 

For Interventionism, anthropocentricity about CAT is only one part of a broader issue. As Pearl himself makes clear, the role of intervention threatens the idea that causation itself is fundamental: 
\begin{quote}
 If you wish to
include the entire universe in the
model, causality disappears because interventions disappear -- the
manipulator and the manipulated
lose their distinction. However,
scientists rarely consider the entirety of the universe as an object of
investigation. In most cases the scientist carves a piece from the universe and proclaims that piece \textit{in} -- namely, the \textit{focus} of investigation.
The rest of the universe is then considered \textit{out} or \textit{background} and is
summarized by what we call \textit{boundary conditions.} This choice of ins
and outs creates asymmetry in the way we look at things, and it is this asymmetry that permits us to talk about ``outside intervention'' and hence about causality and cause–effect
directionality. \cite{Pearl09}   
\end{quote}

In case you feel tempted to respond `So much the worse for Interventionism', look again at {Einstein Locality.} As Einstein says, ``\textit{External} influence on A has no direct influence on B.'' That looks very much like Interventionism. It is doubtful if we can formulate any notion of Locality, or indeed \textit{any causal notions at all,} without implicitly relying on intervention. It is built into assumptions about what we treat as exogenous variables. 

But perhaps we can at least do without an anthropocentric temporal arrow? After all, doesn't physics permit a time-symmetric notion of intervention? At least in a deterministic framework, we might interpret Einstein's `external influence' in terms of an imagined change to properties in a small region of a Cauchy surface at an intermediate time. Such a change, propagated forwards and backwards in accordance with the relevant dynamical laws, will `produce' changes elsewhere.\footnote{Here's a more homely example I once used elsewhere. Consider the perspective of someone
planning to remake the entire series of Star Wars movies, with some tweaks to central characters. The prequels are required
to be consistent with the original Episode IV, and hence tweaks made there will affect the plots of the remakes in both directions. If we substitute Harry Potter for Luke Skywalker the ramifications will spread backwards as well as forwards in the temporal dimension of
the series.} Intuitively, Einstein Locality would be the requirement that such changes propagate only by continuous processes within the lightcones. (Spacelike influences would be allowed, but only indirectly, by indirect zig zags via the lightcones.) 

This may be a good way to capture the sense of causality that matters to relativity, where the idea that there is some sort of fundamental temporal asymmetry seems entirely gratuitous. It might seem an attractive approach for a retrocausalist, too, but I think it throws out far too much, in two senses. First, working science is simply not like this, either in the cafeteria or the laboratory. Our ordinary notions of causality, in science as in everyday life, are those of agents embedded in time with a particular temporal orientation. Throwing all that away would leave the view hostage to the objection rightly raised against CPH, that it is incompatible with the assumptions we need to do science. (As I explained above, CFH is immune from that objection.)

Even more seriously, this symmetric option simply obscures the subtle idea at the core of the retrocausalist proposal. This idea is that \textit{even by the lights of the ordinary asymmetric perspective,} it is possible that the world contains an indirect kind of retrocausality -- hard to notice cases in which by intervening \textit{in the future,} we can make a difference \textit{in the past.} 

Summing up, I want to make three points. First, nothing we know about CAT justifies taking it as fundamental, in the sense in which Bell's Local Causality differs from Einstein Locality. Second, it is doubtful whether Locality, or indeed any interesting notion of causality, can be captured qualitatively in wholly fundamental terms. The right response to this is not to abandon talk of causality in physics. Instead we should keep a close eye on the role of the agent's perspective, in order to keep in mind a question like this: What sort of fundamental structure looks like \textit{this} from \textit{here?} (Travis, I think we are on the same page in wanting such a story.)

Third, coming back to something I mentioned briefly above, we're going to need a distinction between \textit{direct} and \textit{indirect} spacelike influence. Einstein Locality rules out the direct kind, but not apparently the indirect kind. This was clear to the pioneer of the approach, Olivier Costa de Beauregard, who pointed out a decade before Bell's Theorem that zig zag causality, via the past lightcones, provided a potential loophole in the EPR argument \cite{Costa53}. It offered spacelike influence, without action at a distance. This distinction is still missed in some quarters. The following example is from an email I received from an experimentalist known for work in confirming the Bell correlations:
 \begin{quote}
    For me, [the] Costa de Beauregard zig zag in space time, which you seem to consider equivalent to retrocausation, is nothing else than nonlocality. The addition of one time-like vector to the past and one time-like vector to the future, connecting the detections, results in a space-like vector, and a causal relation between both ends, spacelike separated,  amounts to a non local relation.
\end{quote}

As I said, Costa de Beauregard originally thought of his idea as a challenge to EPR. The EPR argument assumes an intuitive notion of Locality, in arguing that measurement choices at A cannot affect measurement outcomes at a remote location B. Costa de Beauregard's point was that if we allow causal influence \textit{in both directions} within the lightcones, and adapt our notion of Locality accordingly, this argument no longer works. There is now a zig zag path for \textit{local} causal influence to reach from A to B. 

So in answer to your question, Travis, Costa de Beauregard's zig zag is still the endpoint that I have in mind. As you rightly point out, it involves rejecting Bell's version of Local Causality. But for the reasons I've sketched, the crucial thing that we need to drop -- that is, CAT, the causal arrow of time -- is on shaky grounds anyway, as a principle for fundamental physics. And Einstein's formulation of Locality from 1948 looks like the alternative that we need.\footnote{I once met Costa de Beauregard, late in his life. I asked him when he had first had the idea for the zig zag, which he first published in 1953. He said in the late 1940s, when he had been a student of de Broglie; but that de Broglie wouldn't let him publish it, until they saw Feynman's work treating positrons as electrons zig-zagging backwards in time.} 


\section{Norsen (II)}

There is a lot going on in that response, all of it very helpful in moving us toward what I see as a possible way of making sense of Bell's perhaps-puzzling dismissiveness about ``fatalism.''  It may take some time to connect the various threads, though!

So, Huw, your vision involves rejecting both \emph{Statistical Independence} and Bell’s \emph{Local Causality}.  But you would hope to preserve a different, time-symmetric notion of locality in which the causal influences on a given event cannot be at space-like separation from it, but are equally allowed to be in either the past- or the future-light-cone.  You suggest calling this alternative notion ``Einstein Locality'' and suggest that it is well-captured by a passage from Einstein’s 1948 essay.  For the record, and despite my appreciation of your praise of the book you quoted the passage from, I have some pretty serious reservations about basing the particular time-symmetric notion of locality that you have in mind on that particular (by the way, translated) passage, and thus attributing your proposed locality concept to Einstein.  But a debate about exactly how to parse Einstein's words will be a pointless distraction here since you've made abundantly clear what you have in mind.  So, having noted my reservation, I will simply follow you in describing your proposed time-symmetric alternative to Bell’s \emph{Local Causality} as ``Einstein Locality.''

My main concern (other than the terminology!) about ``Einstein Locality'' is that, as you acknowledge, ``we're going to need a distinction between \emph{direct} and \emph{indirect} space-like influences.'' From your point of view, the whole purpose of Einstein Locality is to endorse, as compatible with ``the sense of causality that matters to relativity'', the zig-zag sort of multi-step, ``indirect'' causal influence across space-like separation that Bell’s \emph{Local Causality} prohibits.  Without this distinction between ``direct'' and ``indirect'', we would be stuck saying that Einstein Locality allows the very thing that it poses as prohibiting (namely, causal influence across space-like separation).  In that case, Einstein Locality wouldn't actually prohibit anything, i.e., it would be rather empty and pointless.

As I think we agree, it thus seems that your proposal of replacing Bell’s \emph{Local Causality} with Einstein Locality swims or sinks with the project of situating this distinction between ``direct'' and ``indirect'' influences in the context of fundamental physics.  And my initial gut reaction is that this project seems rather hopeless.  When I look at extant theories that possess the appropriate sort of professionalism (e.g., Maxwellian electrodynamics, general relativity, some non-orthodox version of quantum theory) I see causal influences taking the form of continuous propagation.  There is nothing ``atomic'', for example, about the sequence of steps whereby one charged particle affects the motion of another nearby charged particle via the intermediary electric and magnetic field.  There are, if you like, a continuous infinity of infinitesimal intermediating steps, but in my opinion that description should be understood as a human theorist’s perspective on something that is, in reality, a seamless whole.  

So is one charged particle exerting force on another nearby charged particle a ``direct'' influence (because the process is a seamless whole) or an ``indirect'' influence (because it can be viewed as consisting of an infinite number of infinitesimal sub-steps)?  I think the only good answer is to reject the question and whatever line of thinking motivated us to pose it.

Huw, I hope you will correct me if I’m wrong, but I get the impression that instead of trying to find a sharp distinction between direct and indirect influences in fundamental physics, you perhaps want to ground that distinction in an appeal to the concept of causation itself, and in particular the notion of ``intervention'', without which, you said, we probably cannot formulate ``any notion of Locality, or indeed any causal notions at all''.\footnote{Incidentally, isn’t Bell’s \emph{Local Causality} a counterexample there?  It doesn’t explicitly involve, and doesn’t appear to me to implicitly rely on, the idea of “intervention”.}  It may well be true that if we restrict our use of cause-and-effect terminology to processes involving agent-intervention, it might provide a clean way to say, for example, that the zig-zag influence depicted in Figure 3, from the setting $a$ to the space-like separated outcome $B$ (via the particle pair state $\lambda$), is unambiguously indirect.  Both the setting $a$ and the pair state $\lambda$ are after all (at least in part) exogenous:  somebody sets the setting, and somebody sets up the equipment in a certain way to produce particle pairs in a certain state (or a certain distribution of possible states), and both of those interventions are inputs to (not subjects of) the model in question there.  In short, we have, in this case, two distinct interventions, which (on this view) implies the two distinct causal influences symbolized, in Figure 3, with the two distinct black arrows. Maybe this could be said to render the effect of $a$ on the space-like separated $B$ unambiguously indirect.

But to me this kind of proposal for grounding the distinction between direct and indirect causal influences also seems highly suspicious and implausible.  We want, at the end of the day, an account of fundamental physics that isn’t afflicted by anything like a ``measurement problem'', i.e., we want to avoid the use of, or need for, anthropocentric or otherwise ``unprofessionally vague and ambiguous'' concepts and distinctions at the fundamental level.  We want, as you quote Pearl as saying, ultimately ``to include the entire universe in the model''.  And, when we do that, I think it is exactly right that ``interventions disappear''.  

Of course, Pearl goes on to suggest that, in practice, ``scientists rarely consider the entirety of the universe as an object of investigation''; that models always (or almost always) include some parts of the world only as boundary conditions; and that ``this ... permits us to talk about `outside intervention'.''  That may well all be true, but I would nevertheless find a formulation of Einstein Locality which required explicit reference to ``intervention'' (to ground the distinction between direct and indirect influences) to be squarely in the ``unprofessionally vague and ambiguous'' category.

Incidentally, it might surprise some readers to know that Bell, who of course invented and applied the concept of \emph{Local Causality}, was in some sense very sympathetic to at least part of the view that Huw quoted Pearl describing as ``Russell’s enigma'', i.e., the idea that, at the level of fundamental physics, causality (at least described as such) is nowhere to be found.  After presenting his formulation of \emph{Local Causality} in ``La Nouvelle Cuisine'', for example, Bell remarks:

\begin{quote}
Note, by the way, that our definition of locally causal theories, although motivated by talk of `cause' and `effect', does not in the end explicitly involve these rather vague notions. \cite[240]{Bell04}
\end{quote}
This, however, does not mean that (appropriately ``professional'') theories do not describe processes which can legitimately be characterized in causal language.  Bell once wrote that, in 
``pursuing [his] profession of theoretical physics'' he was required to ``insist ... on the distinction between analyzing various physical theories, on the one hand, and philosophizing about the unique real world on the other hand'' \cite[101]{Bell04}  Continuing:

\begin{quote}
In this matter of causality it is a great inconvenience that the real world is given to us once only.  We cannot know what would have happened if something had been different.  We cannot repeat an experiment changing just one variable; the hands of the clock will have moved, and the moons of Jupiter.  Physical theories are more amenable in this respect.  We can \emph{calculate} the consequences of changing free element in a theory, be they only initial conditions, and so can explore the causal structure of the theory. \cite[101]{Bell04}
\end{quote}
This comment occurred in the context of explaining that his concept of \emph{Local Causality} (and, for example, the conditional probabilities which appear in its formulation) should be understood as referring to theories (which are, in turn, candidate descriptions of the unique real world) rather than to the unique real world directly.  Physical theories (at least the serious ones that aspire to fundamentality) thus not only, in Bell’s view, have ``causal structures'', they are our best and necessary tool for, in the long term, discovering the causal structure of the real world. 

Anyway, let me summarize my concern about ``Einstein Locality''.  I fear that, unlike Bell’s \emph{Local Causality}, this notion will never be formulatable in a meaningful and appropriately fundamental way that accomplishes, Huw, what you want it to accomplish.  I fear, in particular, that it will be impossible to cleanly distinguish ``direct'' from ``indirect'' influences in an appropriately ``professional'' manner, and I fear that, without the terminological check provided by an appeal to explicit ``interventions'', an Einstein Local theory of the sort you claim to want will be riddled through with the sort of causal influences across space-like separation that (however you might want to classify them) are just \emph{prima facie} contrary to ``the sense of causality that matters to relativity''.  (On this last point, I am in complete agreement with your experimentalist correspondent.)  

In the same discussion with Shimony, Clauser, and Horne that has been referenced several times already, Bell wrote, about the idea of saving \emph{Local Causality} by rejecting  \emph{Statistical Independence} via the CPH:
\begin{quote}
A theory may appear in which such conspiracies inevitably occur, and these conspiracies may then seem more digestible than the nonlocalities of other theories.  When that theory is announced I will not refuse to listen, either on methodological or other grounds.  But I will not myself try to make such a theory. \cite[103]{Bell04}
\end{quote}
I feel this same way about the idea of rejecting both \emph{Local Causality} and \emph{Statistical Independence} but preserving ``the sense of causality that matters to relativity'' with some notion of Einstein Locality.  I doubt this could be done, and so am not interested in spending my own time and effort on the project, but would be delighted to listen if and when somebody puts forward a precise formulation of Einstein Locality and/or a concrete example of a candidate fundamental theory which shows, if only in principle, how this project could work.

Am I correct, Huw, that neither exists at present?  I have to admit that I was somewhat confused by your proposal to formulate Einstein Locality ``in terms of an imagined change to properties in a small region of a Cauchy surface at an intermediate time.''  That actually sounded rather promising to me, so I was puzzled that in the end you seemed to reject it as ``throw[ing] out far too much.''  

In particular, I didn’t understand the worry that this was somehow ``incompatible with the assumptions we need to do science''.  It seems to me that this formulation is perfectly compatible with ordinary scientific practice.  Indeed, don't several existing (serious candidate) theories, e.g., Maxwell’s electrodynamics, respect this condition?

It seems to me that your rejection of this formulation -- what I think you expressed when you said that it ``obscures the subtle idea at the core of the retrocausalist proposal'' -- must be based on the recognition that such theories as Maxwellian electrodynamics do not appear to support the specific sort of zig-zag causality that would allow for a non-conspiratorial violation of \emph{Statistical Independence}.  But then it's not clear to me what would.

Let me try putting all my cards on the table here.  In my response so far, I haven’t really addressed a core aspect of your proposal, namely, the tension between the apparently time-symmetric fundamental laws, and Bell’s time-asymmetric \emph{Local Causality}.  You are clearly committed to the idea that time-symmetry is fundamental; as you explained from the very beginning, this is the motivation for the whole retrocausalist project.  By contrast, I’m more open to the possibility that some kind of \emph{Causal Arrow of Time} (CAT) may remain in our fundamental physics.  For one, I am not comfortable presupposing that the fundamental laws will turn out to be deterministic; maybe, as the founders of quantum mechanics seemed to believe, the world will turn out to be irreducibly stochastic.  Although I never found the arguments of the founders convincing, I don’t think this possibility has been ruled out, and as long as that remains true I think it is premature to insist that the fundamental laws are time-symmetric.  (Of course, it is also not certain that irreducible stochasticity requires time-asymmetry.  But to me at least it doesn’t seem like time-symmetry and irreducible stochasticity play well together.)  But even if the fundamental laws do turn out to be time-symmetric, I am not convinced this means that there can not, or should not, be something like a fundamental causal arrow of time.  

That said, though, such a fundamental CAT, like explicit notions of `cause' and `effect', may not appear as such, may be relatively invisible, in the formulation of a candidate fundamental theory -- or at least a deterministic candidate fundamental theory.  Indeed, in the context of such theories, my concern about the retrocausalist proposal is not so much that backward causation is  impossible or unconscionable, but rather that the distinction between forward and backward causation seems to melt away.  In Maxwellian electrodynamics, for example, the state of the particles and fields at one time determines the state of the particles and fields at a later time.  So did the former cause the latter, or vice versa?  The laws of the theory certainly don’t answer that question; they just tell us that the states at the two times are necessarily connected.  (This, I take it, was Russell’s point.)  So is Maxwellian electrodynamics a retrocausal theory?  Maybe?  I’m honestly not even sure what the question means.

Of course, at the non-fundamental level, where we model only some narrow part of the universe and describe its surroundings as ``exogenous'' variables through which we might ``intervene'' on the narrow part under study, the distinction between forward and backward causation seems much clearer.  If the system changes due to an intervention in its past, that's forward causation, whereas if the system changes due to an intervention in its future, that's backward causation.  But if we demand that the vague and anthropocentric classification of beables as ``endogenous'' or ``exogenous'' -- if we demand that reference to ``intervention'' -- should disappear at the more fundamental level, it just seems like the distinction between forward-in-time and backward-in-time influences will have to disappear too.  

So I tend to think that you retrocausalists fool yourself into thinking there's a meaningful program to pursue here, by taking a much too interventionist perspective on causation, i.e., by thinking too exclusively about very narrow models of specific situations in which various things are treated explicitly as ``exogenous'' ``interventions''.  And in particular I think that as soon as you try to imagine embedding one of these models, e.g., the one pictured in Figure 3, into a candidate fundamental theory, in which everything is treated on an equal footing, the very notion of retrocausation -- and with it the distinction between the CPH and the CFH -- will disappear like a mirage.

Could I be wrong about all of this?  Absolutely.  To me, the easiest and best way to find out would be to scrutinize a concrete example of a serious candidate fundamental theory that respects a time-symmetric Einstein Locality condition and which (unlike Maxwellian electrodynamics?) supports the needed kind of indirect, zig-zag causality in (but only in!?) the EPR-Bell type of setup where the retrocausalist needs \emph{Statistical Independence} to be violated. 

Unfortunately, I don’t think any such theory exists at present, and everything I’ve said here should make pretty clear why this doesn’t surprise me.  But if (or when) I’m wrong, and such a theory is presented, I will not refuse to listen.  

So, yeah, that’s what my cards look like.  If it turns out I’m (at least arguably) not wrong about all of this, there will be at least a bit more to say about how this relates to Bell's apparent conflation of the CPH and the CFH, i.e., his perhaps-puzzling dismissal of the retrocausality program on the grounds of ``fatalism''.  But Huw, I think I should pause here and give you a chance to weigh in on what I’ve just been saying. 

\section{Price (III)}

In my opening section I described my view that Bell's Theorem contains a loud hint from Nature. Bell shows that if we assume Statistical Independence (SI), quantum theory implies Nonlocality. The hint turns on the thought that we should read this as a \textit{reductio ad absurdum,}   and conclude that SI fails in the quantum realm.\footnote{I don't mean absurdity in the logical sense, of course, but rather what Newton had in mind, when he said of the the idea that `one body may act upon another at a distance through a vacuum, without the mediation of anything else' that it was `so great an absurdity, that \ldots\ no man who has in philosophical matters a competent faculty of thinking, can ever fall into it.'} This would be the obvious reading if we could already see on the shelf some plausible way in which SI might fail, but we don't. Looking further back on the shelf there are two possibilities, CPH and CFH -- the former evidently much easier to see, from most vantage points. I suggested that deafness to the hint might rest on failing to distinguish them, and hence on the view that Bell's well-founded objections to CPH would apply to any attempt to abandon SI. 

In response, Travis, you pointed out that Bell's own assumption Local Causality would still fail in CFH, and asked in what sense I could therefore claim to be defending Locality (or avoiding Nonlocality, as the reasoning just described requires). I offered Einstein Locality as a substitute for Bell's notion, and you have now said that you doubt whether the  distinction  between  the  two  will  be  expressible  in  vocabulary  permitted  by fundamental theory.

It won't matter if this turns out to be the case, in my view, because the argument can simply fall back on Lorentz Invariance. If we treat violation of Lorentz Invariance, and the need for a preferred frame, as the \textit{absurdum} avoided by giving up SI, then the hint speaks to us just as before. To put this another way, suppose we concede to my famous 2015 correspondent that Costa de Beauregard's zig zag proposal still counts as Nonlocality. No matter, so long as the zig zag offers us a path to an explanation of the Bell correlations that avoids the tension Bell himself saw between quantum theory and special relativity. It has often been noted that there are two elements to the counterintuitive character of Nonlocality in contemporary physics, the first linked to relativity and the avoidance of preferred frames, and the second, obviously much older,\footnote{See Newton's remarks above.} to the counterintuitiveness of action at a distance itself.  
 A restriction to the language of fundamental physics might prevent us from expressing the latter, but presumably not the former. And this will do just fine, for the case for the hint. 

As I said in \S2, I now feel that there are two hints, one from Lorentz Invariance and one from Time Symmetry. Concerning the latter, I want to stress again that I think that quantum theory introduces a new reason, not present in the classical regime, for thinking that time symmetry requires retrocausality. A common objection to the retrocausal proposal is a challenge I could paraphrase like this: 'What about time-symmetric classical physics? Is that retrocausal?' This presents retrocausalists like me with a dilemma. If we say `Yes', we are admitting that retrocausation isn't novel or interesting, because it is common in classical physics; if we say `No',  we have conceded that time symmetry alone doesn't imply retrocausality, leaving it unclear what would do so. 

Travis, I take you to be expressing the latter part of this challenge in remarks such as  this: `[S]uch theories as Maxwellian electrodynamics do not appear to support the specific sort of zig-zag causality that would allow for a non-conspiratorial violation of \emph{Statistical Independence}.  But then it's not clear to me what would.' In response, I want to outline what I now think of as the best case for thinking that quantum theory is different.

\begin{figure}[t!]
 \centering
\includegraphics[height=9cm]{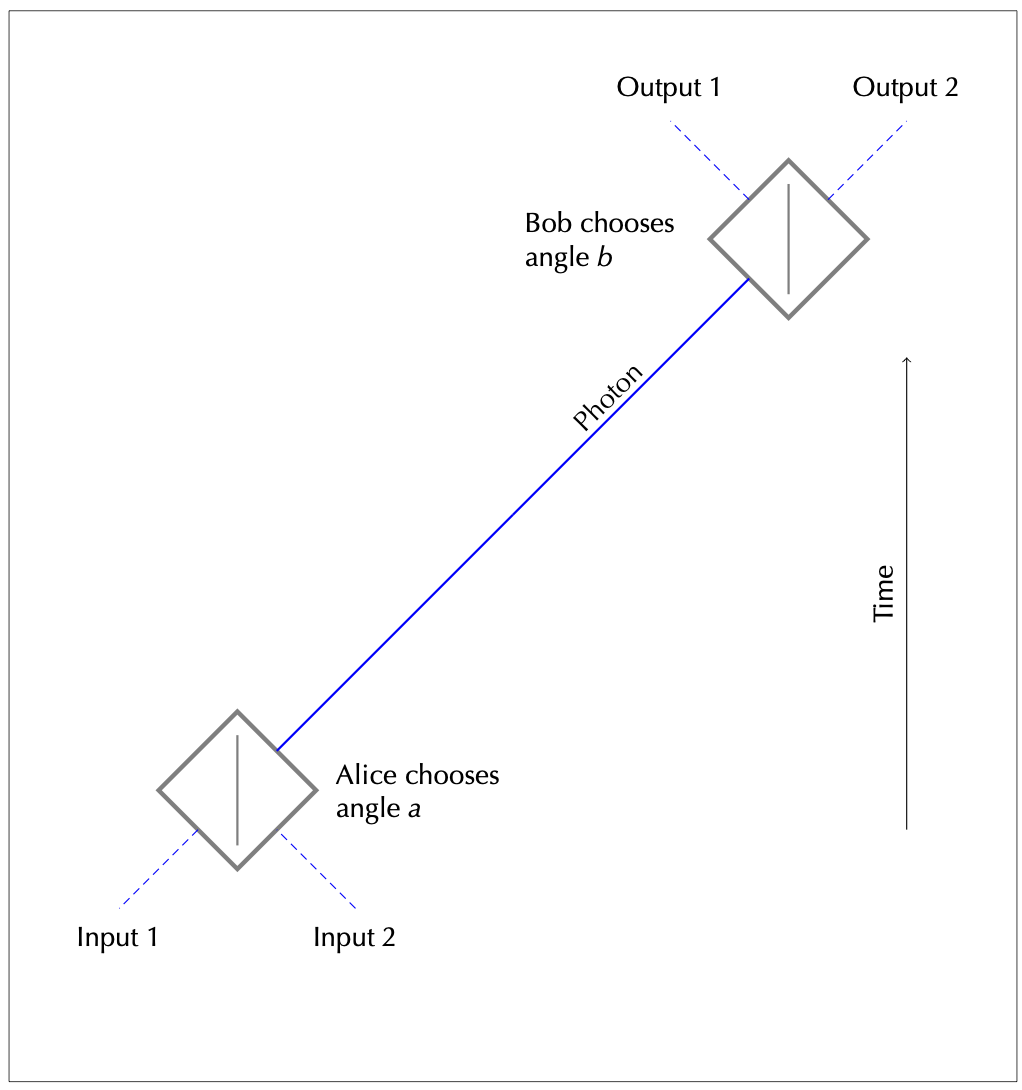}
\caption{\centering\small A timelike EPR-Bell experiment}
\end{figure}

The argument emerged from discussion in \cite{Evans13} of timelike versions of EPR-Bell experiments, such as the one depicted in Figure 4. The `future' end of the experiment involves a polarising beam splitter, as used in many standard spacelike  EPRB experiments. The `past' end involves the same kind of device used in reverse, with a photon entering on one of two channels. In this timelike one-photon experiment the input--output correlations depend on the relation between the settings $\alpha$ and $\beta$ of the two polarisers, just as in a regular spacelike two-photon EPRB experiment.

Normally, the earlier experimenter Alice would be able to control the inputs as well as the setting $\alpha$, and would hence find it easy to signal to Bob, using photon polarisation to carry the required information. The key new idea in \cite{Price12} is that we  restrict Alice's control, to make her situation analogous to Bob's. We do this by putting the input photon under the control of Demons, assumed to know the setting $\alpha$. If the Demons are required to put the input photon on one input channel or the other -- i.e., they are not allowed a superposition of the two -- then their knowledge and options mirror those of Nature at Bob's end of the experiment, in two senses. Nature knows Bob's setting $\beta$, and is required to produce a photon on one output channel or other (at least when a measurement is made). 

Let's call the no superpositions rule the \textit{Discreteness} condition. Without it, the Demons have complete control over the polarisation of the photon between the two devices. Knowing $\alpha$, they can choose weights for an input superposition to produce any polarisation they want.  With Discreteness, however, Alice retains very substantial control. She controls the polarisation completely, up to an additive factor of $0$ or $\pi/2$. (This factor depends on which input channel the Demons choose.) Nevertheless, this degree of control  doesn't guarantee that Alice can signal to Bob. If the Demons choose the input channel at random, and Alice doesn’t know it when she chooses the setting, the control that results from Discreteness does not permit Alice to signal.

So in this artificial situation -- interesting because of the way it mirrors normal circumstances at the other end of the experiment -- {Discreteness introduces what looks by ordinary interventionist lights to be a new element of \textit{forward} causality.} In effect, it gives Alice an \textit{extra} degree of control over the probabilities at Bob's end of the experiment, compared to the case in which the Demon is not restricted in this way.\footnote{As \cite{LeiferPusey17, Leifer17} point out, this is effectively the EPR reasoning, transferred to the timelike case.}   If this extra degree of control is reflected in an underlying ontology, and the ontology is time-symmetric, it will give Bob the same degree of control over the ontology at Alice's end of the experiment. This will amount to a violation of SI and to retrocausality, though not of a kind that would support signalling to the past, for the same reason as in Alice's case. 

The full version of this argument comes with caveats I haven't mentioned here.\footnote{Some of these caveats are removed in the generalisation by  \cite{LeiferPusey17, Leifer17}.}   But I hope I have said enough to explain why I take it on the one hand that violation of SI and retrocausality are not \textit{automatic} consequence of time symmetry; but on the other hand that quantum theory has  features that \textit{may} make them so. Moreover, the subtlety of the new kind of forward causality revealed by this argument offers at least the beginnings of a response to a different challenge: If there were retrocausality, why wouldn't it be obvious and everywhere? The response is that even its forward twin is hard to see.\footnote{Note that in familiar time-asymmetric models, in which the intermediate polarisation depends on Alice's setting but not Bob's setting, the forward causality explains the correlations all by itself -- no retrocausality needed. This suggests that in time-symmetric models, with causal influence in both directions, the forward and backward components could be even more subtle, because they share the explanatory work.} In effect, we had to move ordinary forward control out of the way first, and then focus on what remains.

Let me come back to the remark of yours I quoted above: `[T]heories as Maxwellian electrodynamics do not appear to support the specific sort of zig-zag causality that would allow for a non-conspiratorial violation of \emph{Statistical Independence}.  But then it's not clear to me what would.' The answer suggested by the reasoning just described is that under the constraint of time symmetry, non-conspiratorial violation of SI emerges from the same place as the Bell correlations themselves. We saw that if we consider Bell correlations in timelike settings, the EPR and Bell arguments reveal a distinctive kind of forward causality. From that point, modulo the assumption mentioned about an ontological basis for this causality, insisting on time-symmetry gets us to violation of SI.  (If we map this back to the spacelike cases, we then have the zig zag causal structure we wanted.)

Travis, I appreciate that you are less convinced than I am about both time symmetry and CAT. As you say, you want to hold open the possibility that one or both fail, perhaps independently. For me, both these options fall in the `I won't refuse to listen' box. But what matters for present purposes is not our differing credences about these things. We seem to agree that -- contrary to what many seem to think -- a time-asymmetric CAT doesn't have the status of a well-established piece of physical or metaphysical lore, something that can simply be invoked without argument, in order to dismiss the hint. On the contrary, work outside physics makes a significant case that the causal asymmetry is \textit{not} fundamental -- which means that we have at least some reason to be suspicious of proposals within physics that assume otherwise, implicitly or explicitly. Similarly, work inside physics gives us enough reason to take time symmetry seriously to imply that we certainly can’t take its failure for granted. So again, there's no sign here of something sufficiently well-grounded to dismiss the hint.

The issue you raise of what goes into fundamental theory is very interesting, and in my view quite difficult. It is hard to make sure that we have eradicated the conceptual traces of our human perspective, especially our asymmetric temporal perspective. We are predictive creatures, always acting for the future on the basis of limited knowledge of the past. It is no surprise at all if we describe the world in terms appropriate for such a viewpoint. But we need to keep an open mind to the possibility that the fundamental level need not be described in these terms. (Again, I think we agree on the principle, even if we have different credences about where it might lead us.)

These lessons are as important for retrocausal approaches as for anyone else. For one thing, there's a risk  that in pursuit of time symmetry, retrocausal approaches  find themselves doubling-up ontology that would be better discarded. At the fundamental level we should try to prune away elements that reflect our time-asymmetric viewpoint, not balance them by adding elements reflecting a time-reversed viewpoint. In my view the so-called Two State Vector approach may be guilty of this mistake \cite{Aharonov08}. 

With these `meta' issues about fundamental theory still open, and the path to CFH invisible to most in foundations of physics, let alone in physics more broadly, I think it is no surprise that we retrocausalists don't yet have anything that could claim to be a `serious candidate fundamental theory', as you put it. I recommend  \cite{FriedrichEvans19} and \cite{WhartonArgaman20} for recent surveys of various approaches that have been proposed. 

In the spirit of putting cards on the table, I'll take this opportunity to record a couple of preferences. 
First, I think that the de Broglie-Bohm  theory (dBB)  provides an attractive and under-explored framework for this approach. It has the advantage of an ontology both clear and sparse -- I'm thinking especially of the sense in which position is the only fundamental property, other properties being contextual and relational. This sparsity reduces the risk that we unwittingly build our own epistemic viewpoint into what is intended to be fundamental ontology. Another virtue is that in dBB probabilities emerge much as in classical statistical mechanics, from a distribution over initial conditions. We don't need fundamental time-asymmetric chances, or anything of that kind.\footnote{I think there's no deep difficulty in the fact that we normally consider a distribution over \textit{initial} conditions; final conditions would do the job just as well.}
Finally, the dBB ontology offers an obvious place to hide some subtle retrocausality -- in fact, two places, namely the particle positions and the pilot wave itself, if we give the latter an appropriate ontic status.\footnote{There has already been some work exploring retrocausal versions of dBB: see, e.g.,  \cite{Sutherland06, Goldstein03}). See also \cite{Leifer17} for the sense in which orthodox dBB is time-asymmetric.}  

Secondly, and independently, I like proposals that seek to show how non-conspiratorial SI-violating correlations might emerge from global constraints -- e.g., recent work by Wharton \cite{Wharton15}, Palmer \cite{Palmer15} and Adlam \cite{Adlam18,Adlam20}. The spirit of this approach is nicely captured by Adlam's remark that `God doesn't play dice, he plays Sudoku'  (\cite{Becker18}).\footnote{In Adlam's version the approach is also linked to an interest in fundamental ontology in what seems to me a very interesting way.}

Let me close by coming back to the idea that if we stick to the language of fundamental physics, it will be impossible to draw my distinction between CPH and CFH. If I understand you correctly, Travis, you think this may explain Bell's apparent wish to lump the two together. I'm thinking of remarks such as this:
\begin{quote}
Indeed, in the context of such [fundamental] theories, my concern about the retrocausalist proposal is not so much that backward causation is  impossible or unconscionable, but rather that the distinction between forward and backward causation seems to melt away.  In Maxwellian electrodynamics, for example, the state of the particles and fields at one time determines the state of the particles and fields at a later time.  So did the former cause the latter, or vice versa?  The laws of the theory certainly don’t answer that question; they just tell us that the states at the two times are necessarily connected.  (This, I take it, was Russell’s point.)  So is Maxwellian electrodynamics a retrocausal theory?  Maybe?  I’m honestly not even sure what the question means.
\end{quote}
In response, I want to distinguish  two questions. First, does causal language appear in the fundamental theory? I think we agree in saying `No' to this, at least if we set aside your lingering attachment to a fundamental CAT. 

Second, does fundamental theory make distinctions which, \textit{when viewed from the ordinary perspective of agents like us,} map onto the distinctions we make in causal language (including those involved in distinguishing between CPH and CFH, or between forward and retrocausal models)? 
Here I say `Yes'. Why? Well, consider the corresponding questions about colour. Are red and green categories in fundamental theory? Obviously not. Do we expect fundamental theory to mark the difference between red things and green things? In some sense, obviously, yes -- that's what it is to take colour to \textit{supervene} on fundamental physics. 

If there's a difference between the colour case and the causation case, it is that causal concepts are much more deeply embedded in scientific practice. Some philosophers would see in this an argument for thinking that causation itself needs to be fundamental, but set that aside. Here we are considering the possibility that causation is not fundamental, or at any rate not part of fundamental physics. My point is that this gives us no reason to reject the supervenience of causation on physics. We should expect causal distinctions to depend on lower-level differences, just as we expect for colour.\footnote{This is not to deny that for both colour and causation, there is also an anthropcentric element to the story about the relation between the higher-level categories and underlying physics.} 

This means that I see no reason to think that fundamental physics won't continue to provide the distinctions we need to make causal judgements, including those needed to distinguish between CPH and CFH. But even if I turned out to be wrong, I think it would leave the case for retaining SI no better off. If we throw out the vocabulary we need to distinguish CPH and CFH, we also throw out the terminology we need to raise Bell's objections to rejecting SI. Even if we set aside such explicitly anthropocentric terms such as `free will' and `fatalism', the remaining objection turns on the idea that science treats measurement settings as exogenous variables. You are imagining that even this notion `should disappear at the more fundamental level'. But if we don't have that notion in our vocabulary, we can't make Bell's objection; and if we do have it, then we can distinguish CPH from CFH.\footnote{Even worse, we saw that the case for CFH (non-conspiratorial violation of SI) runs very close to the EPR and Bell arguments themselves. If it really were true that fundamental theory prevented us from making the former, I think it would follow that we couldn't discern the latter. So there's a danger that by blocking the hint we would deprive ourselves of the vocabulary to describe the problem itself.}

\section{Norsen (III)}

In response to my skepticism about (so-called) Einstein Locality, you suggest that we ``can simply fall back on Lorentz Invariance.''  But Lorentz Invariance is nearly as ambiguous and problematic as the various notions of locality we’ve been discussing.  For example, theories with something like a ``preferred frame'' (which idea you mentioned, I think, as something obviously incompatible with the spirit of fundamental relativity) can be Lorentz Invariant and indeed can be argued to be fundamentally relativistic in a serious sense. \cite{hbd,rbm}
I also personally have questions about what compatibility with relativity could or should mean for theories (like virtually all extant, serious, and empirically viable quantum theories) which postulate non-local beables such as the wave function.  

But even leaving that issue aside, I remain confused by the same big-picture point I raised initially.  We know that it is possible to violate Bell’s \emph{Local Causality} in a Lorentz Invariant theory, in several distinct ways. \cite{Goldstein03,hbd,rgrwf}
So if the goal is just to reconcile Bell’s theorem with Lorentz Invariance, we can do this with (Lorentz Invariant) non-locality while still maintaining \emph{Statistical Independence}.  What is the motivation for instead rejecting both \emph{Local Causality} and \emph{Statistical Independence}, by introducing retro-causation?  Of course, I cannot rule out the possibility that going down that road will yield a theory that, despite in some sense rejecting more of Bell’s assumptions than is minimally necessary, is more natural or believable overall as a way of reconciling Bell’s theorem and the associated experiments with fundamental relativity.  But, as I have said, I would personally want to be more convinced that the idea of retro-causality was even coherent, in the context of a candidate fundamental theory without ``measurement problem'' issues, before investing my own time and energy on such a project.

On that issue of coherence, I must admit that I remain somewhat unsure about how you'd answer the question -- ``What about time-symmetric classical physics?  Is that retrocausal?'' -- that you explained poses a dilemma for the retrocausalist.  I think your answer must be no, because otherwise I wouldn’t understand why you’d bother with the somewhat complicated setup and associated argument purporting to establish ``that quantum theory is different'' -- different, I gather you mean, from straightforwardly time-symmetric theories (such as classical mechanics) in supposedly possessing the somewhat subtle and previously-unrecognized sort of causal (including specifically retro-causal) influences that you discuss.

Unfortunately, though, I find this argument totally unconvincing.  It is based completely on the orthodox/textbook version of quantum mechanics including in particular the collapse postulate whose presence is the very core of the measurement problem.  The crucial assumption you call \emph{Discreteness} is just a kind of time-symmetrized collapse postulate, and, as you acknowledge in the more detailed discussion of Ref.~\cite{Price12}, the argument fails to work for the two extant candidate quantum theories with time-symmetric dynamics, namely the de Broglie – Bohm (dBB) and Everett theories.\footnote{I will also note that the alternative notion of time-symmetry developed, for example, in Ref.~\cite{LeiferPusey17}, as a generalization of your argument, is in my opinion revealed to be inappropriate/irrelevant by the fact that manifestly time-symmetric theories such as dBB and Everett do not respect it.}  I would summarize the situation by saying that the argument fails to work \emph{precisely because} these theories eliminate the need for a collapse postulate, i.e., because they do not suffer from the measurement problem.

Yes, as you point out there, this doesn't mean that such theories necessarily exclude retro-causality.  Indeed, as I indicated before, the fundamentally time-symmetric character of the dynamical laws in those theories makes me perfectly open to admitting that they have retro-causality to exactly the same degree or extent, whatever that is exactly, that they have regular forward-causality.  But I gather, from the fact that, if I'm understanding correctly, you suggest somehow changing or supplementing dBB -- ``to hide some subtle retrocausality'' -- that you don’t see any retro-causality, of the sort you want and need, in that theory's standard extant formaultion.  But I don’t understand that proposal at all.  In Ref.~\cite{Price12} you suggest that, in order to deny that dBB is retro-causal in the needed sense, ``it needs to be assumed that \emph{neither} the wave function \emph{nor} the initial positions of the particles are affected by later measurement choices.''  But both elements of the dBB ontology, the wave function and the particle positions, just obey deterministic evolution equations.  Aside from ``external fields'', which would obviously not be included when the theory is applied to the world as a whole, there is simply no room for outside influences, by ``measurement choices'' or anything else.  

To me, that is, it appears that the idea that dBB is somehow a promising candidate for retro-causality of the sort you want and need, arises only from a failure to appreciate that dBB can and should be thought of as a candidate \emph{fundamental} theory, which does not need (and indeed does not even \emph{allow}) special \emph{ad hoc} exceptions to the basic dynamical postulates, associated with measurement or anything else.  To me it instead seems clear that the only sort of retro-causality one could plausibly attribute to the de Broglie – Bohm theory is just exactly the sort that one could, with equal plausibility, attribute to Everettian quantum theory, Maxwellian electrodynamics, or classical particle mechanics.  But, again, and as far as I can tell, this is a sort of retro-causation that you acknowledge is not what you need, so you don’t even call it retro-causality.  

To summarize that point, it seems to me that in order to find the sort of thing you're looking for, you retrocausalists need to set aside the best extant candidate fundamental time-symmetric quantum theories (dBB and Everett) and instead work with a time-symmetrized version of a theory that suffers from ``measurement problem'' issues (or, as in Ref. ~\cite{LeiferPusey17}, implausibly re-define the meaning of ``time-symmetric'').  This just reinforces my previously-expressed sense that the very concept of retro-causation, of the sort you want and need, is a kind of mirage.  It appears to be meaningful only when you put interventionist causality in by hand, by backing away from the fundamental level of description and instead treating certain things as ``exogenous'', i.e., outside of the quantum system under study and amenable to some kind of agent control.  Doing this is, admittedly, part of the essential character of orthodox quantum theory.  But that doesn't make it right. 

You suggest that although causal language will likely not appear, as such, at the fundamental level, a candidate fundamental theory should probably still ``make distinctions which, \emph{when viewed from the ordinary perspective of agents like us}, map onto the distinctions we make in causal language.''  I completely agree.  Causality should supervene on fundamental physics even if it doesn't appear there explicitly labeled as such.  And as I've said I'm completely open to the possibility that, for example, in the context of a candidate deterministic time-symmetric fundamental theory like dBB, there might well be some legitimate grounds for speaking in terms of retro-causal influences.  I don't exactly see what those grounds might be, but I'm happy to leave the door open.  But at least as long as the fundamental physics remains deterministic, it seems to me that any correlations of the sort needed to violate \emph{Statistical Independence} -- even ones that we end up agreeing make sense to describe, ``when viewed from the ordinary perspective of agents like us'' who are part of the world described by the theory, in terms of retro-causality and the CFH -- will imply correlated correlations, so to speak, in the physical state of the world at some much earlier time. 

Let me try to explain more clearly what I'm trying to get at here.  There is some equipment in the lab that is arranged, in a certain way, to produce a sequence of particle pairs whose state we describe with the variable $\lambda$.  And the experimenters set up some pieces of measuring equipment, including, say, some random number generators, which perform, on the incoming particle pairs, measurements of the particles’ spins along directions $a$ and $b$.  We know, I take it, that $\lambda$ is causally influenced by (even if not completely determined by) the state of the lab equipment:  if something is not plugged in, no particle pairs will emerge at all; if some optical element is mis-aligned, the pairs may emerge in (what QM would describe as) a triplet state instead of the intended singlet state; etc.  And similarly, the precise sequence of settings, $a$ and $b$, is in fact determined by various details of the setup.

To make things really concrete, let's suppose that the particle source produces pairs with states coming in a certain order, starting from the moment the equipment was most recently powered up.  And suppose that morning there was a windstorm in town, which resulted in a tree falling on a power line, briefly interrupting the supply of electricity to the lab, and thus causing the particle source to reboot and reinitiate its sequence of emitted pairs at that particular moment.  And, similarly, suppose that the seed for the pseudo-random number generating algorithm was chosen, on this occasion, by the number of blueberries in the lab assistant's pancake at the diner earlier that morning.  

Now, in a fully deterministic theory like dBB, there does not seem to me to be any basis for claiming, nor does there appear to be any room for adding, causal influences from the settings, $a$ and $b$, onto $\lambda$.  The settings are just determined by various things in their past, as are the pair states.  And so to posit a \emph{Statistical-Independence}-violating correlation between these is to posit a very special, ``just so'' type of correlation between the precise physical details of the morning windstorm which resulted in that particular tree being blown down at that particular moment, and the precise physical details of the factors determining the subtle movements of the diner chef's hand which resulted in, say, eleven, rather than ten or twelve, blueberries ending up in that one particular pancake.  This is the sort of correlation needed to violate \emph{Statistical Independence} with the CPH, and I think we agree that it seems unacceptably conspiratorial.

The new point I want to make here is that I do not see, exactly, how violating \emph{Statistical Independence} instead with the CFH leads to a less conspiratorial picture.  Suppose I keep an open mind and allow for the possibility that some hypothetical future theory might include, in some meaningful and compelling way, retro-causal influences from the settings $a$ and $b$ onto the pair states $\lambda$.  My point here is that, presumably, that theory will also have to acknowledge the causal connection (and hence tight correlation) between the precise sequence of settings $a$ and $b$ and whatever complicated factors determined how many blueberries ended up in the lab assistant's pancakes.  And it will also, presumably, have to acknowledge the causal influence of the morning windstorm on $\lambda$.  I don’t know quite how the distinct causal influences on $\lambda$ -- one (backwards in time) from $a$ and $b$, and one (forward in time) from the morning windstorm -- would be reconciled in such a theory.  Maybe the theory would say that $a$ and $b$ are, alone, sufficient to \emph{determine} $\lambda$, and would thus be forced into saying that $\lambda$, instead of being influenced by the morning windstorm, retro-causally determines the relevant details of that morning windstorm.  Or maybe there really would be two oppositely-directly causal influences on $\lambda$, neither of which alone would be sufficient to \emph{determine} lambda, but which, together, are.  In this scenario as well, we would (presumably?) still have to end up with a very tight correlation between the morning windstorm and the settings, and therefore also between the morning windstorm and the blueberries.  

This is precisely the sort of correlation, between seemingly-random details of the states of seemingly-unrelated phenomena in the past, which we regard as unacceptably conspiratorial in the context of the CPH.  Why should such correlations be considered any less conspiratorial in the context of the CFH?  For this reason, to me, the distinction between violating \emph{Statistical Independence} with the CPH, and violating it with the CFH, becomes blurry at best when we zoom out, include more of the world in our system, and refrain from treating various elements in a special way, as ``exogenous'' ``interventions''.  

You claimed, at the end of your most recent contribution, that to whatever extent we lose the ability to make the CPH/CFH distinction, we will also lose, to that same extent, the ability to argue for the reasonableness of \emph{Statistical Independence} (or even to rehearse the EPR argument and Bell’s theorem) in the first place.  I don’t understand this.  As discussed in the exchange with Shimony, Clauser, and Horne that Bell referenced in his letter to you, the main case for \emph{Statistical Independence} is a practical and empirical one.  In particular, \emph{Statistical Independence} is assumed in virtually every scientific experiment.  Think, for example, of a randomized controlled drug trial whose standard interpretation requires assuming that the (say) coin flips, determining which patients got the drug and which the placebo, were uncorrelated with the previous health of the patients:  if all of the patients who got the real drug survive and all the patients who got the placebo die, we would ordinarily infer that the drug works well to prevent death.  But of course it's possible that the drug has no positive effect at all; instead, the coin came up tails for, and hence the placebo was given to, all and only those patients who had some other health condition, totally unrelated to that targeted by the drug, and were destined to die of that.  As Shimony, Horne, and Clauser put it, in a passage you quoted earlier but which bears repeating here, denying Statistical Independence 
\begin{quote}
...will essentially dismiss all results of scientific experimentation.  Unless we proceed under the assumption that hidden conspiracies of this sort do not occur, we have abandoned in advance the whole enterprise of discovering the laws of nature by experimentation.\cite{Bell85}
\end{quote}
Science, that is, relies on the \emph{Statistical Independence} assumption, and science unquestionably works.

This practical case for \emph{Statistical Independence} does not appear to me to rely on any interventionist perspective on the applicability of causal terminology to fundamental theories.  Indeed, it is easy enough to understand what \emph{Statistical Independence} means in the context of such theories.  For the randomized drug trial, it means that the specific facts, in the distant past, which determine the precise sequence of coin flips (used to assign placebo or drug to each patient) should not be conspiratorially correlated with the specific facts which determine whether those individual patients will or won’t die of some unrelated malady.  Similarly, in the EPR-Bell case, \emph{Statistical Independence} means that the specific facts which determine the sequence of measurement settings $a$ and $b$ (e.g., the blueberries) should not be conspiratorially correlated with the specific facts which influence the sequence of particle pair states $\lambda$ (e.g., the windstorm). 

Of course, Huw, you’ll want to say I'm begging the question here by inserting the word ``conspiratorially''.  Your whole point is that, if the settings retro-causally influence the pair states, the needed correlations between (on the one hand) $a$ and $b$ and (on the other hand) $\lambda$ would not need to be conspiratorial at all.  

By way of concluding, let me try to summarize the several reasons I have explored, throughout this dialogue, for being skeptical of this project.

First, I am skeptical that, at the level of a fundamental candidate theory (i.e., without ``interventions'' and ``exogenous'' variables) the idea of retro-causation, of the rather special sort you want and need, even makes sense.  (When I look at extant candidate fundamental theories, at least the time-symmetric deterministic ones we have available, the idea of assigning specific temporally-oriented causal arrows to specific sub-processes seems arbitrary and groundless.  At the fundamental level, the theories just assert necessary connections between states at different times.)  

Second, I am skeptical of the claim that, if you did somehow produce a candidate fundamental theory that somehow made your desired sort of violation of \emph{SI} compelling, the theory would respect some meaningful ``Einstein Locality'' or Lorentz Invariance condition.  (It seems more likely to me that, without the terminological restraints introduced by interventionism, your ``indirect'' zig-zag causality would end up being ubiquitous and the theory would end up being wantonly non-local and blatantly incompatible with relativity.)

 The new, third, grounds for skepticism that I have been trying to raise here has to do with the fact that, even in a (deterministic) candidate fundamental theory in which it somehow made sense to speak of the settings $a$ and $b$ retro-causally influencing the pair states $\lambda$, something still determines the settings, and some other factors, besides $a$ and $b$, will presumably still have to have some influence on $\lambda$.  And these influences, traced further backwards in time, will imply some very particular and special correlations, in the earlier states, which, at least from this abstract perspective, do not seem different from --  and so do not seem any less conspiratorial than -- the correlations posited by the CPH.  

A concrete candidate theory could, in principle, cleanly refute any or all of the grounds for skepticism I've expressed here.  But so far I have never seen a candidate retro-causal theory with the appropriate ``pretension to physical precision''. I’ve instead only seen various sorts of toy models that focus on some narrow system, \emph{postulate} ``interventions'' by treating certain variables as ``exogenous'', and (hence) suffer from a kind of ``measurement problem''.  I have not, that is, seen anything that refutes my skepticism, so I remain unconvinced that the retro-causal program generally, and the distinction between the CPH and CFH in particular, can survive the translation to a more serious, candidate fundamental theory.

Was it similar reasoning that led Bell, in the letter reproduced in Fig 1, to dismiss your retro-causal project?  I'm not sure.  If anything, it seems more likely to me that he just didn’t appreciate that you wanted to violate \emph{Statistical Independence} in a non-standard and purportedly non-conspiratorial way.   But I think it's at least possible that he, as a consistent champion of the need for fundamental theories which avoid the measurement problem, recognized that, from this fundamental point of view, the standard and non-standard ways of violating \emph{SI} -- that is, the CPH and CFH -- are very difficult to even distinguish.  

That, at any rate, is where I, having been profoundly influenced by Bell, end up.  At least for the time being.

\section{Price (IV)}

Travis, it's a little frustrating to close at this point, when your latest comments raise several points for further discussion. But wrap up we must, so I’ll confine these closing comments to just one issue, that of `conspiracy'. Before that, I want to say a very warm thanks to you for taking this on. As you know, the idea for this dialogue originated in an exchange of Facebook comments \cite{Facebook10}, after I shared my letter from Bell there. Some fascinating discussion takes place on Facebook these days, but it has been a pleasure, as well as an education, to do this the old, slow way.\footnote{`Too swiftly now the Hours take flight! /
What's read at morn is dead at night', as Austin Dobson -- a man well ahead of his time! -- noted already in the 1880s \cite[233]{Dobson89}.}  

Now to conspiracies. Helpfully, you suggest an answer to your own objection: 
\begin{quote}
[Y]ou’ll want to say I'm begging the question here by inserting the word ``conspiratorially''.  Your whole point is that, if the settings retro-causally influence the pair states, the needed correlations between (on the one hand) $a$ and $b$ and (on the other hand) $\lambda$ would not need to be conspiratorial at all.  
\end{quote}
That would do, but I want to say a bit more. For one thing, I want to make sure it's clear why I take  CFH to be much less vulnerable to the conspiracies charge in the first place. 

I introduced the term `conspiracy' into our discussion in \S2. It occurs there in the quote from Shimony, Clauser and Horne \cite{Bell85} that you quote again. But I didn't use the term  myself in distinguishing CPH and CFH. For me, the crucial difference was simply that unlike CPH, CFH treats measurement settings in the normal way, as exogenous variables. I noted that this explains how CFH escapes the objection that these authors and many since have raised for CPH, that it is incompatible with `the whole enterprise of
discovering the laws of nature by experimentation'. However, a few paragraphs later I added that I hadn't dealt with the suggestion that CFH also requires something `conspiratorial'. Let me now come back to that, and explain why it is a very different issue from the one that confronts CPH.

In your account of the conspiracies objection, you note that physicists (notoriously playful folk) can easily put measurement settings under the control of such things as the number of blueberries in a pancake. And you point out -- rightly, if you are talking about CPH -- that this means that any theory trying to break \textit{Statistical Independence} is going to have to concern itself with blueberries. Just look at the two (fortuitously) blue arrows in Figure 2. If there are blueberries on the causal chains that those arrows represent, the common cause needs to control the blueberries, along with everything else in the chain. Quite an ask!

But now look at Figure 3, depicting CFH. Here, the guts of the retrocausal proposal is some sort of  lawlike constraint, correlating the measurement setting $a$ with properties $\lambda$ of the lefthand particle along the first black arrow. (The direction of the arrow just reflects the fact that $a$ is an exogenous variable, on which interventions are possible.) Once we have these guts, the story about the blueberries follows for free. Blueberries are simply one among the endless ways that ingenious experimenters can devise to control the value of an exogenous variable (i.e., to provide the green arrow in Figure 3). We don't need anything novel or conspiratorial to control \textit{them.} 
By way of comparison, imagine someone puzzled about how the buttons on a remote handset control a television. There's no additional mystery about how blueberries, too (and everything in the past on which they themselves depend), can control the television, if we hook them up to  the handset.\footnote{To take this analogy a little further, we could imagine blueberries being used as a source of (effectively) random experimental interventions, in a project to test the hypothesis that settings of the handset have a causal influence on the television. In that project it would be absurd, of course, to assume statistical independence between the handset settings and the state of the television -- what we would be looking for would be, precisely, \textit{violations} of such independence. This trivial example shows that science doesn't assume such statistical independence \textit{everywhere,} and that it is indeed begging the question against CFH to assume it in the form of the principle we've been calling \textit{Statistical Independence.}}  

So CFH is far less vulnerable to the charge of conspiracy than CPH. 
If there is something seemingly conspiratorial in CFH, it is the fact that the particles `already know' the measurement settings, in some sense, before they arrive at the measurement devices. This certainly looks strange to ordinary intuitions, and words such as `conspiratorial', or `teleological', do something to capture this strangeness.  

If someone objects to CFH on this basis I'm inclined to accuse them of a temporal double standard -- in other words, as I used the term in \cite{Price96}, applying different principles in one direction of time to the other, without offering any justification for the difference. This kind of `conspiratorial' behaviour is exceedingly common in the direction from future to past, thanks to the thermodynamic asymmetry. Think of all the time-reversed videos you have ever seen of omelettes transforming themselves into unbroken eggs, and similar things. Many writers take that kind of behaviour to be explained by a lawlike constraint in the past, the so-called Past Hypothesis. 

I am not suggesting that CFH involves a time-reversed version of the familiar thermodynamic asymmetry. But if anyone objects to this remaining sense in which CFH looks conspiratorial, I will ask them whether they would still object to the time-reversed version of the same theory, in which a particle and a measuring device are correlated in virtue of a lawlike constraint on an interaction in their common past. If not, I'll accuse them of a temporal double standard. That's close to the charge of begging the question, but a bit more specific in a useful way.\footnote{I discuss these issues at length in \cite[Ch.~5]{Price96}.}

Obviously there's more to say here. For one thing, I haven't said anything about your concern that CFH requires $\lambda$ to be controlled from two directions. Indeed, reading through the whole thing again, it feels like we're just warming up. But for the moment, thanks again, Travis -- and I hope this won't be our last opportunity to discuss these questions. Perhaps this dialogue  will encourage others to weigh in on one side or other, and take up the open issues.   

\section{Norsen (IV)}

I share the sense that we are only now getting warmed up.  But I think, rather than be frustrated to wrap up just when it feels like we're finally getting to the heart of the issues, we should appreciate the progress that getting to this point represents.  And of course the end of this particular dialogue is not the end of dialogue as such.  Our illuminating and highly enjoyable discussion here can hopefully generate and influence further constructive discussions in the future (and, not to rudely ignore the possibility of retro-causation, perhaps also in the past).  

By way of wrapping up, I will just touch briefly on two points, making a special effort not to unfairly lob any new rhetorical grenades.  In particular, I want to flag, for the purposes of future discussion, a point where I think we may have a substantive and unresolved disagreement.  And then I will attempt to clarify my (admittedly potentially misleading) use of the word ``conspiracy'' in my previous contribution.

So, first, the flag.  You said that my argument ``that any theory trying to break \emph{Statistical Independence} is going to have to concern itself with blueberries'' would apply ``if you are talking about CPH''.  I think you meant to imply that this argument does not apply in the context of CFH.  But the whole point of my admittedly silly parable involving blueberries and windstorms, is precisely that it should still apply, either way.  

You wrote that, for you, ``the crucial difference [between CPH and CFH] was simply that unlike CPH, CFH treats measurement settings in the normal way, as exogenous variables.''  But perhaps the fundamental theme running through everything I've written here is that everything we might want to say about the EPR-Bell scenario needs be compatible with the perspective of a candidate fundamental theory which is free of ``measurement problem'' issues and in which, in particular, there \emph{are} no exogenous variables.  

So if the reason you think that the CFH is \emph{not} ``going to have to concern itself with blueberries'' is that the CFH \emph{by definition} treats measurement settings as exogenous (and is hence freed from the responsibility of considering the factors that, in a real-world implementation of this setup, would in fact influence/determine the settings) then I would regard that as a fatal flaw in the CFH program.

As always, though, I suspect there may be some mutual misunderstanding that makes our differences appear greater than they really are.  In particular, your reference back to the (indeed, fortuitously) blue arrows in Figure 2 gives me pause.  You say:  ``If there are blueberries on the causal chains that those arrows represent, the common cause needs to control the blueberries, along with everything else in the chain.  Quite an ask!''

But I believe the way you are thinking of \emph{Statistical Indepedence} being violated here, in the context of the CPH, is not what I had in mind.  It is of course true that \emph{Statistical Independence} could easily be violated if, instead of the particle pair source spitting out pairs in a pre-defined order determined by the precise moment and location of the windstorm, the pair source instead spits out a sequence of particles in states that are, say, determined by the output of the very same random number generator (seeded by the blueberry count) that is also determining the settings.  This sort of arrangement would seem to be the sort of thing you had in mind (in which the blueberries, the pair states $\lambda$, and the settings $a$ and $b$ are all on the same future-directed causal chain.  But there would be nothing remotely conspiratorial about the failure of \emph{Statistical Independence} in this kind of case.  We would just blame the failure on an exceedingly stupid experimental design by the experimenters.

What I had in mind -- what I thought we agreed would count as unacceptably conspiratorial --  is instead the possibility that \emph{Statistical Independence} could be violated (still leaving aside the possibility of retro-causation, i.e., working still in the framework of the CPH) with a more sensible experimental design, in which the causal chain leading up to the pair states $\lambda$ and the chain leading up to the settings $a$ and $b$, have no apparent connection.  That was the point of the parable, with the windstorm affecting the one thing and the (causally disconnected) blueberries affecting the other.  To me, the interesting question is whether \emph{Statistical Independence} might still be violated in this kind of case.  And of course it might be.  But its being violated (with the non-stupid experimental design) would require certain subtle details in the kitchen of the diner to be correlated, just so, with certain subtle details of the weather pattern across town.  The usual attitude, though, is that there is no reason those details should be correlated.  So positing that they are -- that is, rejecting \emph{Statistical Independence} -- amounts to asserting something that has a highly implausible, ``conspiratorial'' feel to it.

So much for the flag.  Now to the clarification.

The point I was making previously is that, by considering the CFH (in which the settings $a$ and $b$ retro-causally influence the pair states $\lambda$), we in no way remove the correlation (between those subtle details in the diner and other subtle details in the weather pattern across town) that we would be committed to with the CPH.  I think in the end you agreed with this; at least, that’s how I understand your comment that ``the story about the blueberries follows for free''.  (But you see, I'm confused, because you also previously insisted that the CFH requires taking the settings as exogenous.)  Assuming that's right, though, I think you then just want to object to my characterization of the correlations as ``conspiratorial''.  They would be (as I thought we agreed) in the CPH, because in that context there is no reason for them.  But, I think you want to say, these same correlations are not at all ``conspiratorial'' in the CFH because the causal connection between the future end of the one chain (the settings) and the future end of the other chain (the pair states) connects the two otherwise-unconnected chains, and thus provides a perfectly good reason for the subtle details (on the past ends of the two, now-connected, chains) to be correlated.  Indeed, as I think you want to say, once we allow retro-causation, those correlations are no weirder, no more conspiratorial, than the future correlations which we already allow must exist between subtle details of things which have interacted in the past.  

If you accept this as a fair summary of your view, Huw, I'm happy to just concede all of it.  I shouldn't have continued to use the inflammatory word ``conspiratorial'' to describe the correlations (between subtle details of things like blueberries and windstorms in the past) that both CPH and CFH are, I think, committed to, when discussing those correlations from the point of view of the CFH.  The correlations are the same, whether we adopt the CPH or the CFH, but their conspiratorialness is not the same.  

But my intention wasn't to beg the question.  Partly I developed the parable and pushed this point just because I think it is important to acknowledge that violating \emph{Statistical Independence} implies these sorts of correlations, regardless of how exactly one does it.  (Treating measurement settings as exogenous variables tends to obscure this fact.)  

And just because these sorts of correlations (in so far as they arise via the CFH) should not be described (and subsequently dismissed) as ``conspiratorial'', doesn’t mean they are beyond reproach.  For example, as I think you hinted at, their existence raises questions about how to reconcile the posited retro-causation with the statistical/thermodynamic arrow of time.  And I have to admit that I feel a vague additional sense of un-ease about such correlations, stretching, as they would clearly have to, all the way back to the big bang.  I have a hard time putting my finger on the reason for this queasiness.  Of course, one possibility is that, try as I might, I can’t quite get myself to take retro-causation fully seriously, so, in my gut, I respond just as I would to the conspiratorial character that those same correlations would have in a theory without retro-causation.  

I do think there is more to it than that, however, and what I meant to gesture toward before was the thought that maybe this could explain what Bell wrote in his letter to you.  If ``fatalism'' is the uncomfortable idea that our apparent freedom (or even just apparent randomness) is illusory -- that what we choose to (or just happen to) do has, in fact, been pre-written in a script going all the way back to the big bang -- then isn’t there a kind of time-reversed fatalism inherent in your CFH?  For me at least, and perhaps for Bell, the idea that my apparently free choices (for example, about how to seed the random number generator that controls the settings in a Bell experiment) end up causing incomprehensibly subtle correlations in the early universe, is no less uncomfortable than the idea that incomprehensibly subtle correlations in the early universe are the true causes of my apparently free choices.

But is there more to this discomfort than a lingering bias against retro-causation?  Answering that will, I think, require further reflection, discussion, and debate.  I look forward to that, and I thank you again for inviting me to participate in this.

\vspace{6pt} 



\authorcontributions{The authors contributed equally to this piece.}

\funding{This research received no external funding.}

\acknowledgments{HP thanks Reinhold Bertlmann for advice concerning the publication of his letter from J.~S.~Bell, and Ken Wharton for comments on drafts.}

\conflictsofinterest{The authors declare no conflict of interest.} 


\reftitle{References}

\end{document}